\documentclass[sigconf, nonacm]{acmart}

\usepackage{graphicx}
\usepackage{textcomp}
\usepackage{xcolor}

\usepackage{makecell}

\usepackage{algorithm}
\usepackage[noend]{algpseudocode}

\usepackage{url}
\usepackage{latexsym}
\usepackage{amsmath}
\usepackage{graphicx}
\usepackage[subfigure]{graphfig}
\usepackage{multirow}
\usepackage{multirow, booktabs}
\usepackage{footnote}
\usepackage{CJKutf8}
\usepackage{booktabs}
\usepackage{microtype}
\usepackage{balance}

\usepackage{listings}
\usepackage{xcolor}
\usepackage[font=normal]{caption}

\newcommand\vldbpagestyle{plain} 

\begin{document}

\title{Speech-to-SQL: Towards Speech-driven SQL Query Generation From Natural Language Question}

\author{Yuanfeng Song$^{1, 2}$, Raymond Chi-Wing Wong$^{1}$, Xuefang Zhao$^{2}$, Di Jiang$^{2}$}
\affiliation{
\institution{$^{1}$The Hong Kong University of Science and Technology, Hong Kong, China  $^{2}$AI Group, WeBank Co., Ltd, China}}
\email{{songyf, raywong}@cse.ust.hk}

\renewcommand{\authors}{Yuanfeng Song, Raymond Chi-Wing Wong, Xuefang Zhao, Di Jiang}

\begin{abstract}
Speech-based inputs have been gaining significant momentum with the popularity of smartphones and tablets in our daily lives, since voice is the most easiest and efficient way for human-computer interaction. This paper works towards designing more effective speech-based interfaces to query the structured data in relational databases. 
We first identify a new task named \emph{Speech-to-SQL}, which aims to understand the information conveyed by human speech and directly translate it into structured query language (SQL) statements. A naive solution to this problem can work in a cascaded manner, that is, an automatic speech recognition (ASR) component followed by a text-to-SQL component. However, it requires a high-quality ASR system and also suffers from the error compounding problem between the two components, resulting in limited performance. 

To handle these challenges, we further propose a novel end-to-end neural architecture named \emph{SpeechSQLNet} to directly translate human speech into SQL queries without an external ASR step. SpeechSQLNet has the advantage of making full use of the rich linguistic information presented in speech. 
To the best of our knowledge, this is the first attempt to directly synthesize SQL based on arbitrary natural language questions, rather than a natural language-based version of SQL or its variants with a limited SQL grammar.
To validate the effectiveness of the proposed problem and model, we further construct a dataset named \emph{SpeechQL}, by piggybacking the widely-used text-to-SQL datasets. Extensive experimental evaluations on this dataset show that SpeechSQLNet can directly synthesize high-quality SQL queries from human speech, outperforming various competitive counterparts as well as the cascaded methods in terms of exact match accuracies. 
We expect speech-to-SQL would inspire more research on more effective and efficient human-machine interfaces to lower the barrier of using relational databases. 

\end{abstract}


\maketitle

\pagestyle{\vldbpagestyle}
\begingroup
\renewcommand\thefootnote{}\footnote{\noindent
A preprint.
}\addtocounter{footnote}{-1}\endgroup

\section{Introduction}

Nowadays, the vast majority of data in our lives is stored in relational databases, and an essential tool for accessing these data is through structured query language (SQL) commands \cite{kedar2009database}. However, SQL has a rather complex grammar and long learning curve, and the difficulty of mastering SQL blocks many non-technical users and data analysts to use SQL. 
To facilitate these users to perform data queries, automatically generating SQL queries from natural language (NL) questions, or text-to-SQL, has been extensively studied in natural language processing (NLP) and database communities recently \cite{li2014constructing,sen2020athena,nihalani2011natural,zhang2019editing,li2014nalir,kim2020natural,affolter2019comparative}. 

Compared with the text-based input, it is widely believed that the voice-based interface is a much easier and efficient way for human-computer interaction. 
According to the user study in \cite{shah2020speakql}, the voice-based interface could allow users to compose SQL queries considerably faster by up to 6.7x compared to typing on a text-based interface in a tablet device. 
Meanwhile, with the popularity of smart smartphones and tablets in our daily lives, dozens of voice-based applications have emerged in the market in the past decades, such as voice-based search engines \cite{alateeq2020voxento}, voice-based AI assistants, and chatbots (e.g., Siri, Xiaoice, Google Home, and Cortana) \cite{peng2020understanding} and voice-based databases \cite{shah2020speakql}. 

In parallel, studies already exist for building voice-based interfaces for database systems in the research community \cite{shah2020speakql,shah2019demonstration,utama2017voice,trummer2020demonstrating}. 
For example, SpeakQL implements a voice-based query interface for structured data that enables users to input SQL with speech. 
Utama \textit{et al.} \cite{utama2017voice} designed a system named EchoQuery, which also supports users to query the database with voice.
However, these studies usually restrict the voice as a query to be an NL-based version of SQL or its variants with a limited subset of SQL grammar, and thus, they still require the users to have professional background and skill in SQL language. For example, EchoQuery requires that the basic query request should be in the form like ``\texttt{\textup{What is the \{Aggregation\}\{Columns(s)\} of \{Table(s)\}?}}'', and SpeakQL requires the query to be an exact SQL statement such as ``\texttt{\textup{Select Salary From Employees Where Name Equals John}}''.
As such, none of them achieves translating common flexible speech-based NL questions into SQL queries, a harder yet more valuable task widely believed to be a more efficient and straightforward way of human-machine interaction that lowers the barrier of using relational databases.

\begin{figure}[t!]
    \centering
    \includegraphics[width=0.50\textwidth]{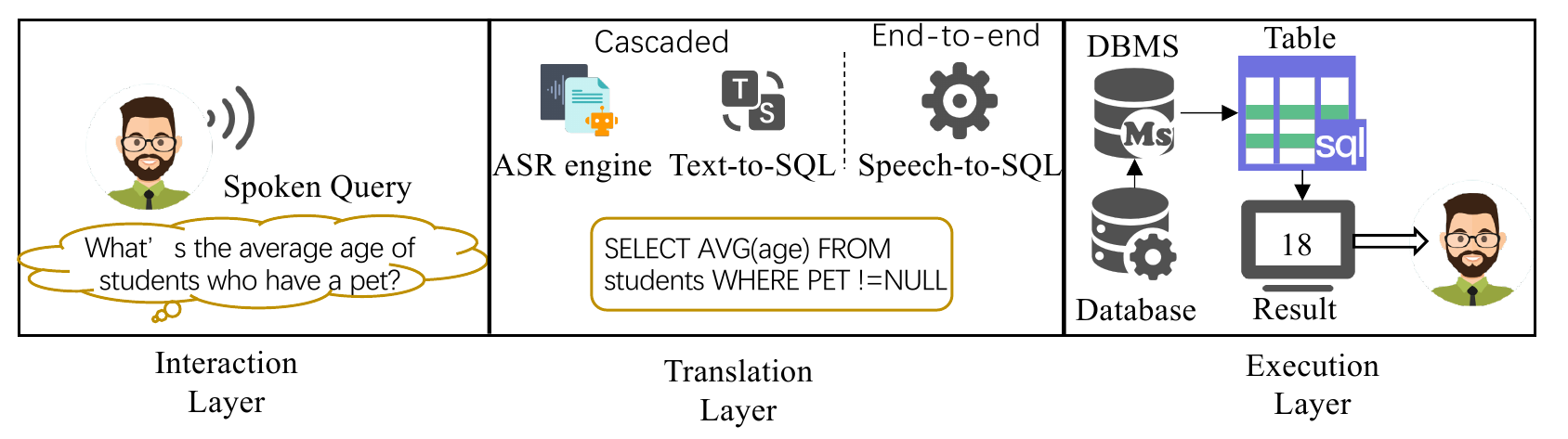}
    \caption{Speech-driven database querying using Speech-to-SQL; 
    Text is shown only for illustrating the speech content, which is not available in real scenarios.}
    \label{fig:task}
\vspace{-12pt}
\end{figure}

In this paper, we work towards designing more effective voice-based interfaces that attempt to handle common NL questions to manipulate the structured data stored in the relational database. Towards this goal, we first present a new task named \texttt{Speech-to-SQL}, which aims to understand the information conveyed by human speech and translate it into the corresponding SQL query, as shown in Figure~\ref{fig:task}. A naive solution to this task can work in a cascaded fashion, namely first converting speech signals into their corresponding transcripts with an automatic speech recognition (ASR) component, and then conducting downstream SQL generation by a text-to-SQL component. 
While this approach can alleviate the problem to a certain extent, it suffers the error compounding problem, that is, the ASR module produces myriad forms of errors in the recognized transcripts which brings a big technical challenge for following text-to-SQL conversion. In our analysis, we found that existing text-to-SQL models are not robust to these ASR errors. For example, according to our analysis, a 33\% ASR error rate would cause more than 36\% accuracy reduction for the downstream SQL conversion. Lastly, from a greater and broader perspective, there are around 7000 languages in the world and nearly half of them lack a commonly used written form \cite{wang2021generating}, which prohibits these languages from benefiting from any existing text-driven NLP technologies. 
This also prompts many parallel studies on {speech-to-X} applications that bypass the text, where ``X'' refers to a wide range of modalities such as text (in a language same as the source one or a translated one) \cite{bansal2017towards}, speech (in a language same as the source one or a translated one) \cite{wahlster2013verbmobil}, image \cite{li2020direct}, general-purpose program code (e.g., Java, Python) \cite{desilets2006voicecode,nowogrodzki2018speaking}, software model \cite{black2019voice}, and so on. 
However, despite these strong motivations, no work has been done on a direct end-to-end speech-to-SQL conversion in literature. 

Motivated by the above-mentioned issues, we aim to explore the end-to-end neural approach that does not require an external ASR step for speech-to-SQL in this work. 
To validate the rationale of this problem and the following possible models, we first construct a benchmark dataset named \texttt{SpeechQL}, by piggybacking the public text-to-SQL datasets - WikiSQL \cite{zhong2017seq2sql} and Spider \cite{yu2018spider}. 
Then, we design an advanced neural architecture, \texttt{SpeechSQLNet}, to directly explore the semantics presented in the speech and synthesize the corresponding SQL queries. 
In particular, {SpeechSQLNet} seamlessly integrates speech encoder, Graphical Neural Network (GNN), and transformer as its backbones to conduct speech parsing for unlabelled speech data without an ASR step as a premise, and the whole network is optimized in an end-to-end fashion. 
The modality gap between speech and text is the biggest challenge for training our {SpeechSQLNet} architecture. 
As such, we design two novel pre-training mechanisms, \emph{speech-sentence} and \emph{speech-item} pre-training, to further boost the performance. 
The speech-sentence pre-training is based on an auto-encoder to force the two modalities to map into the same hidden space. 
The speech-item pre-training aims to identify the references of tables and columns in the speech queries.
Experimental evaluations on this proposed dataset show that {SpeechSQLNet} could synthesize high-quality SQL queries directly from speech. 
We hope speech-to-SQL will become a popular task and show some insights into more effective and intelligible speech-driven human-machine interfaces for relational databases.

The contributions of the work can be summarized as follows.

$\bullet$  We propose a kind of new query interface with its corresponding task, {Speech-to-SQL}, that lowers the barriers of using SQL and relational databases.
To promote further research in this task, we construct a benchmark dataset {SpeechQL}. To the best of our knowledge, this work is the first attempt in the literature that aims to directly synthesize SQL from an arbitrary human question in speech and bypasses text. (Section~\ref{sec:pro})

$\bullet$  We explore two approaches, the cascaded one and the end-to-end one (i.e., {SpeechSQLNet}), to synthesize the SQL queries from speech. 
Compared with the cascaded one, SpeechSQLNet has the advantage of reducing the error-compounding problem by optimizing in an end-to-end style, achieving a better performance compared with the cascaded methods. (Section~\ref{sec:naive} \& Section~\ref{sec:model})  

$\bullet$ To bridge the challenging modality gap between speech and text, we propose a novel speech-sentence pre-training mechanism that employs an autoencoder-based framework to map the speech and text into the same hidden space. Furthermore, to handle the schema-linking problem, we propose another speech-item pre-training mechanism to empower the modal's capability in identifying the refereed items in the speech query. (Section~\ref{subsec:pre-training})

$\bullet$  We conducted extensive experiments on this proposed benchmark dataset, compared with several strong baselines. Experimental results show that {SpeechSQLNet} can significantly outperform not only other end-to-end baselines but also improve the cascaded models, such as an advanced model - IRNet \cite{guo2019towards}, by up to 11.96\% in terms of exact match accuracy. (Section~\ref{sec:exp})

The rest of this paper is organized as follows. We first introduce some preliminaries in Section~\ref{sec:pre}. 
Then we give the detailed problem setup in Section~\ref{sec:pro}, following by the introduction of the naive cascaded approach in Section~\ref{sec:naive}.
The proposed SpeechSQLNet, including the model structure as well as its components, is illustrated in Section~\ref{sec:model}.
The specially-designed training mechanism is discussed in Section~\ref{subsec:pre-training}. 
Experimental evaluations are then presented in Section~\ref{sec:exp}, followed by a comprehensive review of the related work in Section~\ref{sec:related}. 
Finally, we conclude the work in Section~\ref{sec:con}.

\section{Preliminaries}
\label{sec:pre}

We introduce some preliminary knowledge about sequence-to-sequence and transformer structure, which helps the understanding of the following concept in this paper. 

\vspace{-5pt}
\subsection{Sequence-to-Sequence Structure}

Sequence-to-Sequence architecture \cite{cho2014learning}, or Seq2Seq for short, refers to a series of neural architectures that map a given sequence of elements (i.e., words, speech signals) into another sequence. Many tasks such as machine translation or ASR enjoy a common Seq2Seq structure and can be solved under the Seq2Seq framework. 
Take English-to-Chinese machine translation as an example, the input sequence of a Seq2Seq network designed for this task is the source language (i.e., English) while the output sequence is the target language (i.e., Chinese). 
The network usually consists of an encoder and a decoder, where the former takes the input sequence and converts it into some hidden representations, and the latter maps these hidden representations into the target sequence. The common choice of the encoder and decoder can be a Recurrent Neural Network (RNN) \cite{medsker2001recurrent} or Long Short-Term Memory (LSTM). The speech-to-SQL problem also enjoys a Seq2Seq structure and can also be considered to be a special case of Seq2Seq conversion. As such, we also employed several basic Seq2Seq network architectures as baselines for performance comparison in this work. 

\subsection{Transformer-based Structure}
\label{subsec:transformer}
As a special case of the Seq2Seq structure, Ashish \textit{et al.} \cite{vaswani2017attention} propose a neural network named transformer that shows promising performance in various NLP tasks such as machine translation and dialogue systems. 
A novel architecture named \textit{multi-head self-attention} is designed, which is formally defined as
\begin{equation}
    \texttt{\textup{MultiHead}}(Q,K,V)=\texttt{\textup{Concat}}(h_1,h_2, \cdots, h_n)W^{o}, 
\label{equ:att}
\end{equation}
where $Q$ is the query, $K$ is the key, $V$ is the value, and $W^{o}$ is learnable parameters. $n$ is the number of heads, and each $h_{i}$ is obtained from an attention module, defines as
\begin{equation}
    h_{i}=\texttt{\textup{Attention}}(QW_{i}^{Q}, KW_{i}^{K}, VW_{i}^{V}),
\label{equ:att}
\end{equation}
\begin{equation}
    \texttt{\textup{Attention}}(Q_i,K_i,V_i)=\texttt{\textup{softmax}}(\frac{Q_{i}K_i^{T}}{\sqrt{d_{k}}})V_i,
\label{equ:att}
\end{equation}
where $d_{k} = d_{\textup{model}} / n $ is the dimension of queries $Q_i$, 
$W_{i}^{Q} \in \mathbb{R}^{d_{\textup{model}} \times d_k}$, $W_{i}^{K} \in \mathbb{R}^{d_{\textup{model}} \times d_k}$, $W_{i}^{V} \in \mathbb{R}^{d_{\textup{model}} \times d_k}$, and $ W^{o} \in \mathbb{R}^{d_{\textup{model}} \times d_{\textup{model}}}$ are the parameters of the network.

The transformer consists of multiple stacked encoder and decoder layers. 
The encoder contains several layers consisting of 
a self-attention module followed by a position-wise feed-forward layer, defined as 
\begin{equation}
\texttt{\textup{FFN}}(X) = \texttt{\textup{max}}(0, XW_1 + b_1)W_2 + b_2,
\end{equation}
where $\texttt{FFN}(\cdot)$ refers to feed-forward network, $W_1$ and $W_2$ are the weight matrices of each layer, $b_1$ and $b_2$ are the corresponding bias, and $X$ is the input matrix.
To further capture the sequential information, a positional encoding (PE) \cite{vaswani2017attention} mechanism is further employed,
which mathematically defined as 
\begin{equation}
\begin{aligned}
\texttt{PE}_{(pos, 2i)}=\sin (pos/10000^{2i/d_{\textup{model}}}),
\end{aligned}
\end{equation}
\begin{equation}
\begin{aligned}
\texttt{PE}_{(pos, 2i+1)}=\cos (pos/10000^{2i/d_{\textup{model}}}),
\end{aligned}
\end{equation}
where $pos$ is the token position in a sequence, $i$ is the dimension index.

\section{Problem Setup}
\label{sec:pro}

After introducing the preliminaries, we are ready to give the formal definition of the speech-to-SQL problem, and the benchmark dataset constructed to evaluate the rationale of the proposed problem and possible models. 
\subsection{Speech-to-SQL Problem}

Suppose that we have a speech corpus $\mathcal{D}$ of $M$ instances, denoted as $\mathcal{D} = \{ \mathbf{d}^{1}, \cdots, \mathbf{d}^{M} \}$, where $\mathbf{d}^{i}$ ($i \in \{1, \cdots, M\}$) represents the $i$-th instance. The superscript $i$ is ignored for simplicity in the following discussion.
Each training instance $\mathbf{d}$ is in the format of $\{{x}, {y}, {S}\}$, where ${x}$ is a speech recording expressing the NL question, ${y}$ is its translation in SQL query, and ${S}$ is the schema of the corresponding database with which ${y}$ will be executed. 
The \emph{speech-to-SQL} problem aims to learn a model which can translate an unseen question-schema pair $\{{x}', {S}'\}$ to its corresponding SQL query ${y}'$. 
Specifically, the schema ${S}$ includes the collection of tables $T_{x}=\{t_i\}_{i=1}^{N}$, the collection of columns $C_{x}=\{c_{i,j}\}_{j=1}^{L_{i}}$ for each table $t_{i} \in T_{x}$, where $N$ is the total number of tables, and $L_{i}$ is the number of columns in table $t_i$. 

\begin{figure}[t!]
    \centering
    \includegraphics[width=0.50\textwidth]{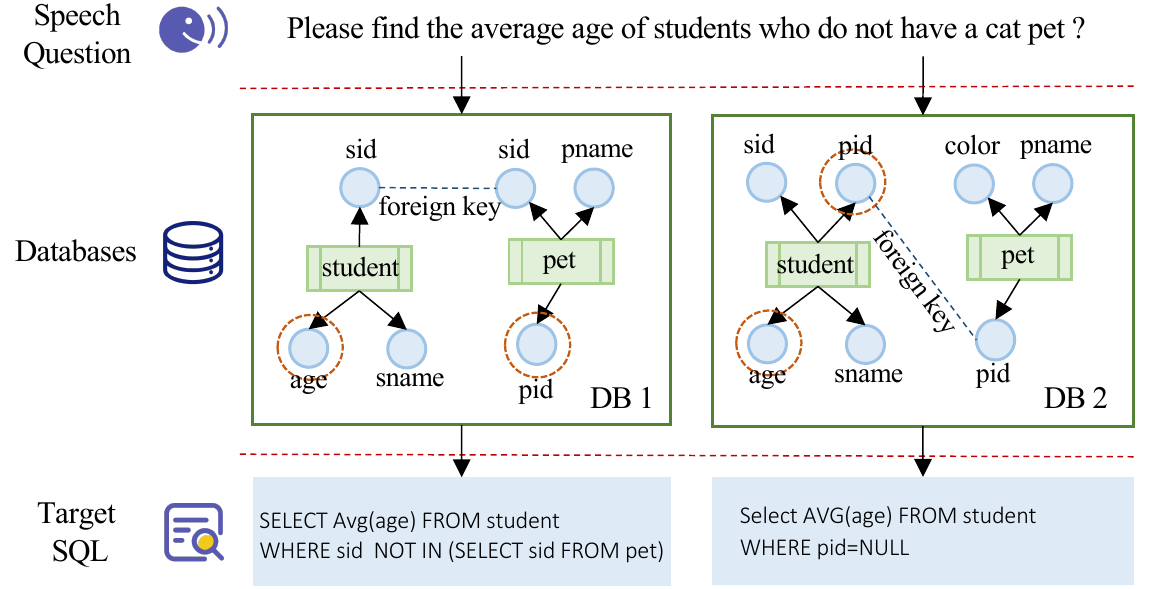}
    \caption{Two examples of the speech-to-SQL problem, showing that different databases (schemas) can result in different desired SQL queries, even for the same speech question.}
    \label{fig:problem}
\vspace{-15pt}
\end{figure}

Figure~\ref{fig:problem} gives two example for a vivid illustration. In the first example, the user expresses their data requirement by simply asking a question ``\texttt{\textup{Please find the average age of students who do not have a cat pet?}}'' 
to the speech-based interface of the databases. 
Given the schema of the database as ``\texttt{\textup{student (sid, sname, age), pet (pid, sid, pname)}}'' and the above speech (without the corresponding transcript), the speech-to-text task aims to directly synthesize the desired SQL query ``\texttt{\textup{SELECT AVG(age) FROM student WHERE sid NOT IN (SELECT sid FROM pet)}}''. Finally, the DBMS could execute the generated SQL and display the final results to the user. 
For the second case, the same NL question is queried by the user on another database with schema ``\texttt{\textup{student (sid, pid, sname, age), pet (pid, color, pname)}}'', and the desired SQL would be ``\texttt{\textup{SELECT AVG(age) FROM student WHERE pid = NULL}}'' correspondingly, to guarantee the correctness of the final results.

\subsection{Speech-to-SQL Dataset}

Since speech-to-SQL is a new task, there are currently no existing datasets in the literature that are suitable for evaluating its performance. 
Following the same practice of using text-to-speech (TTS) technology \cite{fan2014tts} to generate the spoken captions in the speech-to-image task \cite{wang2020s2igan,wang2021generating,li2020direct}, we create a new dataset based on public text-to-SQL datasets WikiSQL \cite{zhong2017seq2sql} and Spider \cite{yu2018spider} by converting the textual statement into speech using a TTS module. 

There are several text-to-SQL datasets in the text-to-SQL field. Spider dataset is a small-scale dataset for cross-domain text-to-SQL evaluation annotated by human experts.
It contains 8,625 training instances, 1,034 validation instances, and 2,147 test instances. 
One of the reasons that Spider is popular in the text-to-SQL field is that it contains diverse databases, ranging from restaurants, scholars, to academics, without overlaps in the training, development, and test sets, which makes it a good benchmark to conduct cross-domain evaluation.
The WikiSQL dataset is another popular dataset in the text-to-SQL task. Compared with the Spider, WikiSQL only covers simple queries in form of \texttt{\textup{aggregate-where-select}} structure, but the size of the WikiSQL dataset is much larger. 
To combine the merits of these two datasets, we merge them and further augment the data with the method proposed in \cite{yu2018syntaxsqlnet}, and finally, we manually check the validity of the generated samples.
As a first step towards speech-to-SQL, over complicated SQL cases are not included and we leave them for future work. 

For the TTS module, we synthesize the raw waveform from the Mel spectrogram using MelGAN \cite{kumar2019melgan}, with a stable and efficient filter bank – pseudo quadrature mirror filter bank (Pseudo-QMF) \cite{nguyen1994near}. 
We then apply FastPitch \cite{lancucki2020fastpitch} to FastSpeech 2 \cite{ren2020fastspeech} to generate the mel spectrogram from text. 
Finally, we obtain a new dataset which we named \emph{SpeechQL}. 
The detailed statistics of this dataset can be found in Table~\ref{tab:dataset}. 

\begin{table}[t!]
\caption{The Statistics of the proposed {SpeechQL} Dataset}
\centering
  \begin{tabular}{l|cccc}
    \toprule
    \textbf{Statistic} & \textbf{Train} & \textbf{Validation} & \textbf{Test} & \textbf{Total}  \\ 
    \midrule
    No. of Instances & 31431  & 1100   & 2200  & 34731 \\
    No. of Databases & 13199  & 1058  & 2023  & 13601 \\
    Length of Speech &  37.43h &  1.33h &  2.61h &   41.37h \\
    Avg NL Length &  11.85 &   11.97 &  11.83 &  11.85 \\
    Avg SQL Length & 9.1 & 9.05  &  9.03 &   9.09 \\
    Vocab Size & - & -  & -  & 15908 \\
    \bottomrule
  \end{tabular}
\label{tab:dataset}
\vspace{-0pt}
\end{table}

\section{A Naive Baseline: The Cascaded Approach}
\label{sec:naive}
A naive solution to the speech-to-SQL task can work in a cascaded fashion, namely, first converting speech signals into their corresponding transcripts with an ASR system, and then conducting downstream text-to-SQL conversion. 
In this section, we mainly discuss this cascaded solution from its main components: ASR in Section~\ref{subsec:asr} and text-to-SQL in Section~\ref{subsec:tts}. 

\subsection{Automatic Speech Recognition}
\label{subsec:asr}

Mathematically, the hybrid ASR component transcribes the user's voice question into a textual output, which can be expressed as follows:
\begin{equation}
\begin{aligned}
\mathbf{w^{*}} & = \arg \max_{\mathbf{w}} (\log P_{LM}(\mathbf{w})  + \lambda \log P_{AM}(\mathbf{a|w})),
\end{aligned}
\end{equation}
where $\lambda$ is a trade-off parameter. $P_{AM}$ is an acoustic model (AM), evaluating how sounds combine to produce word sequence, and $P_{LM}$ is a language model (LM), picking the word sequence that has the largest probability in human language perspective \cite{song2021l2rs,song2020goldenretriever}.

The ASR component plays a critical role in the cascaded approach for speech-to-SQL, however, constructing an SQL domain ASR is hard since the NL questions contain many domain-specific words, especially in the database contents. 
These words can easily be misrecognized, for example, in a real case from SpeechQL dataset, the original question is ``\texttt{\textup{Which Video has a PSIP Short Name of rt?}}''. It is as easily-recognizable as ``\texttt{\textup{What video has a safe short name of artie}}'', even by the widely-used ASR engines provided by Google \footnote{\url{https://cloud.google.com/speech-to-text}}, Microsoft AI Cloud \footnote{\url{https://azure.microsoft.com/cognitive-services}}, or Baidu Cloud \footnote{\url{https://ai.baidu.com/tech/speech/asr}}.  
The reason for this phenomenon is obvious, that is, the trained data used to construct their ASR model usually lack SQL domain-specific data. 
Hence, it would be better and flexible if we could construct an ASR model by ourselves with a SQL-domain dataset.

To alleviate the above-mentioned problem, we turn to a two-step (i.e., pre-training and fine-tuning) approach to build a reliable and accurate ASR model dedicated to the SQL domain. 
Specifically, we first train the ASR model with a large-scale in-house English dataset with around 8000h speech from general domains, then the pre-trained ASR model is further fine-tuned with the small-scale text-to-SQL data to adapt to the database domain. 
For the ASR system, we utilize the Kaldi ``Chain'' model \cite{povey2011kaldi} as the AM component and employ a trigram LM as the LM component. 
The DNN component of the ``Chain'' model is a Time Delay Neural Network (TDNN) \cite{waibel1989phoneme}. 
To further improve the performance, an advanced rescoring mechanism named L2RS \cite{song2020goldenretriever,song2021l2rs}, integrated with BERT \cite{devlin2019bert}, is also adopted in the $N$-best rescoring step. 
We employ the widely-used metric, Word Error Rate (WER), as the main indicator for the performance of the constructed ASR model. 
WER is formally defined as 
\begin{equation}
  \begin{aligned}
  WER = \frac{S+D+I}{T},
  \end{aligned}
  \end{equation}
where $S$, $D$, and $I$ represent the number of word substitutions, deletions, and insertions, respectively, and $T$ is the total number of words in the ground-truth transcript. From the definition, we can conclude that WER reflects the quality of the ASR decoding result, and the lower the value, the better the decoding result. 
Overall, we achieve a WER of 
34.451\%, a relatively low error rate in this specific SQL domain. 
The well-tuned ASR model makes the cascaded approaches very strong baselines since parallel studies in the speech-to-image task such as \cite{wang2021generating} employ the ASR model with a far high WER (around 50\%) as their baseline.

\subsection{Text-to-SQL Conversion}
\label{subsec:tts}

The text-to-SQL model aims to convert the NL questions into SQL queries, with previous studies such as Seq2SQL \cite{zhong2017seq2sql}, SeqNet \cite{xu2018sqlnet}, EditSQL \cite{zhang2019editing}, IRNET \cite{guo2019towards}, and so on.
In our implementation, we directly construct text-to-SQL models with existing techniques on our datasets. 
The existing text-to-SQL model is usually trained based on a clean dataset, that is, the NL questions contain no errors. 
However, in our scenario, the NL question is recognized by an ASR module, which produces myriad forms of errors in recognized transcriptions that further introduce the error compounding problem, a technical challenge for following text-to-SQL conversion. 
To improve the performance, we then explore the possibility of end-to-end speech-to-SQL conversion. 

\begin{figure*}[t!]
    \centering
    \includegraphics[width=0.9\textwidth]{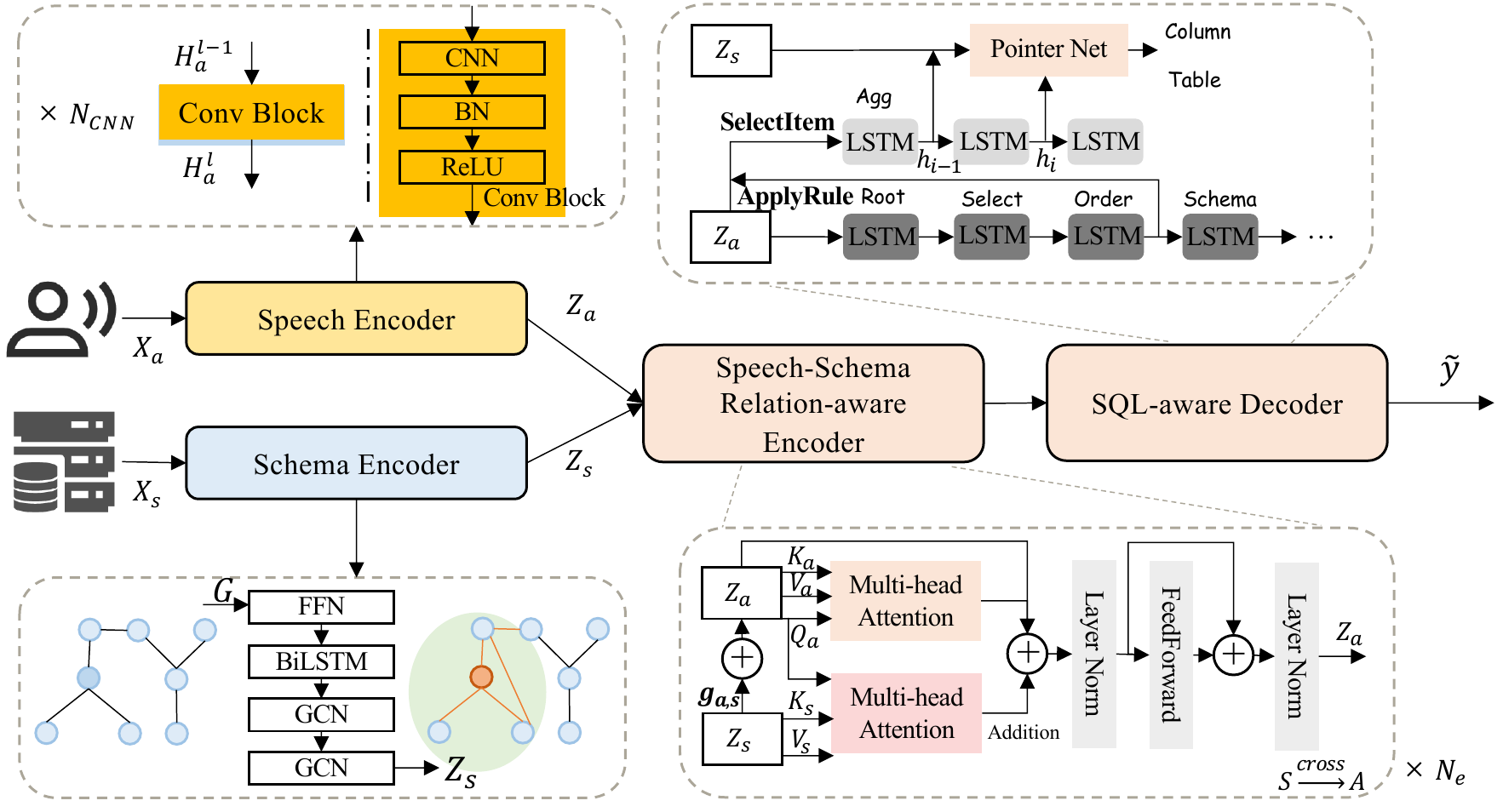}
    \caption{The Network Structure of the Proposed SpeechSQLNet Model, including a speech-encoder, a schema-encoder, and a SQL-aware decoder. The output of SpeechSQLNet is an abstract structure tree $\tilde{y}$ in SemQL \cite{guo2019towards}, which is further converted into the desired SQL query $y$.}
    \label{fig:framework}
\vspace{-10pt}
\end{figure*}

\section{Our Proposed End-to-End Model: SpeechSQLNet}
\label{sec:model}

Directly synthesizing SQL queries from speech signals is hard due to the modality gap between speech signals (i.e., audio modality) and database schema (i.e., text modality) . 
Inspired by the common text-to-SQL architectures \cite{xu2018sqlnet,rat-sql,guo2019towards}, the SpeechSQLNet model first uses a speech encoder to convert the speech signals into hidden representations. In the meantime, the schema, which greatly affects the desired SQL, is also converted into hidden features by a GNN-based encoder to preserve its structural information. Finally, the speech embedding, together with the schema features, is fused to synthesize the
corresponding SQL query with semantic consistency. The overall structure of the proposed model is illustrated in Figure~\ref{fig:framework}.
In this section, we detail the proposed SpeechSQLNet model from its four main components: Speech Encoder (Section~\ref{sec:sub-speech-encoder}), Schema Encoder (Section~\ref{sec:sub-schema-encoder}), Speech-Schema Relationship-aware Encoder (Section~\ref{sec:sub-speech-schema-relation}), and SQL-aware Decoder (Section~\ref{sec:sub-sql-decoder}).

\subsection{Speech Encoder}
\label{sec:sub-speech-encoder}

\subsubsection{Raw Feature Extraction from Speech Signals}
Theoretically, it should be possible to synthesize SQL directly from the digitized waveform, due to the strong modeling capability of exiting DNNs. 
However, human speech signals are highly variable, and the objective of conducting feature extraction is to reduce the variabilities. Specifically, the typical variability that we would like to eliminate includes the effect of the periodicity or pitch, the amplitude of excitation signal and fundamental frequency \cite{chazan2000speech}.
Inspired by the common practice in ASR, in our implementation, we extract 96 log-scaled Mel-band energies from speech $x$ using sliding windows of 1024 samples ($\approx 46 ms$), with 512 overlaps and the Hamming windowing function, to serve as inputs $X_a \in \mathcal{R}^{l_a \times 96}$ for speech encoder.

\subsubsection{Speech Embedding from CNN-based Architecture}
After extracting the feature embeddings from the digitized waveform, we design a speech encoder that employs a convolutional neural network (CNN)-based structure to preserve the continuity of speech. Specifically, the network is composed of a series of convolutional blocks, each of which is stacked by a CNN module \cite{krizhevsky2012imagenet} and a batch normalization (BN) \cite{ioffe2015batch} module. The process of each block can be denoted as:
\begin{equation}
\begin{aligned}
H_a^{(l)}=f(\texttt{BN}^{(l)}(\texttt{CNN}^{(l)}(H_a^{(l-1)}))),
\end{aligned}
\end{equation}
\begin{equation}
\begin{aligned}
H_a^{(0)}=X_a,
\end{aligned}
\end{equation}
where $l= \{1, \dots, N_a\}$ is the layer number of stacked blocks, $H_a^{(l)} \in \mathbb{R}^{l_a^{(l)} \times d_a^{(l)}}$ is the hidden representation of the speech features, delivered from the $l$-th layer and sent to the $(l+1)$-th layer, $l_a^{(l)}$ and $d_a^{(l)}$ are the length and size of $l$-th layer speech hidden representation, respectively. $f(\cdot)$ is a activation function and in our network, we choose the popular $\texttt{ReLU}$ function \cite{glorot2011deep}. 
Lastly, we obtain a speech embedding $Z_a \in \mathbb{R}^{l_a^{(N_a)} \times d_a^{(N_a)}} $ by the speech encoder as 
\begin{equation}
\begin{aligned}
Z_a = H_a^{(N)},
\end{aligned}
\end{equation}
where $N_a$ is the number of layers in speech stacked blocks.

\subsection{Schema Encoder}
\label{sec:sub-schema-encoder}
Even for the same NL question, it has been proven that the schema of the database greatly influences the structure of the desired SQL query \cite{bogin2019representing}. 
As such, we need to preserve the schema information from both the semantic and the structural perspectives. 
In SpeechSQLNet, we first convert the database schemas into graphs, and then employ the GNN-based architecture as the schema encoder. 

\begin{figure}[t!]
    \centering
    \includegraphics[width=0.45\textwidth]{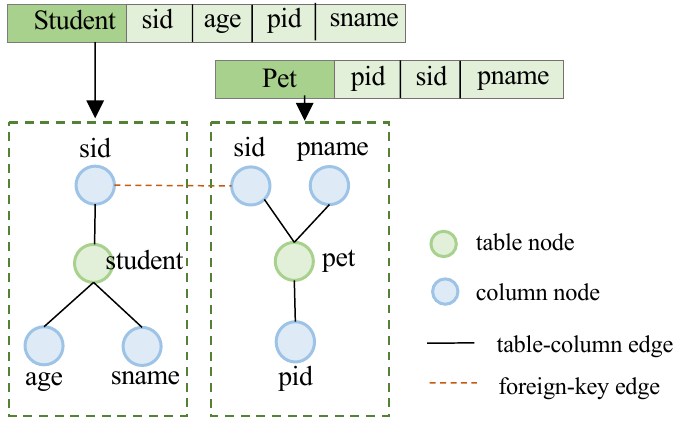}
    \caption{An Example of Converting Database Schema into Graph}
    \label{fig:schema-graph}
\vspace{-10pt}
\end{figure}

\subsubsection{Converting Database Schema into Graph}
To leverage the structure information, we define an undirected graph $G = \{V, E\} $ to represent the schema. 
An example with two tables is shown in Figure~\ref{fig:schema-graph}. 
Specifically, each node $v_{i} \in V$ is either a table $t \in T_{x}$ or a column $c \in C_{x}$, and each edge $e_{i,j} = (v_{i}, v_{j}) \in E$ is designed based on the database relations. 
There are two types of edges in the graph, and each of them represents a way how columns and tables correlate to one another, including table-column (edge between table $t_i \in T_{x}$ and its column $c_{j} \in C_x = \{c_{i,j}\}_{j=1}^{L_i}$) and foreign-key (edge between foreign key linking two different tables).
Considering that items with the same name contribute to the feature construction and SQL decoder in the same way, the columns with the same name, even in a different table, are represented by the same node. Finally, for each node $v$, we could get its neighborhood $N(v) = \{u \in V | {u,v} \in E\}$. 
Node sequences and their adjacent matrix are serving as inputs $X_s$ for the schema encoder.

\subsubsection{Converting the NL into Node Embedding}
To retain the semantic information, instead of randomly initializing the embedding of each node in the graph, we adopt an LSTM-based network to generate the embedding. 
For a node $v_{i} \in V$, each token in the node is first converted into an embedding vector, and then a bidirectional LSTM (BiLSTM) is employed to convert the variable-length node tokens' embedding into fixed-length hidden state vectors. 
The output vectors of the forward and backward LSTM are concatenated together, followed by a feed-forward network (FFN) to construct the node embedding $h_{v_i} \in \mathbb{R}^{d_{model}}$.
Then the initial embedding of the schema graph, denoted as $H_{s} \in \mathbb{R}^{l_{s} \times d_{model}}$ which includes $l_{s}$ nodes, is calculated as follows.
\begin{equation}
\begin{aligned}
h_{v_i}&=\texttt{FFN}(\texttt{BiLSTM}(embed(v_i))),
\end{aligned}
\end{equation}
\begin{equation}
\begin{aligned}
H_{s} &= [h_{v_0},\dots ,h_{v_{l_s}}].
\end{aligned}
\end{equation}

\subsubsection{GCN-based Schema Encoder}
Given the schema in the form of a graph, we would like to capture the overall structure and the detailed connections for each node. 
As such, we use a GNN-based encoder to embed the schema graph $G$ into a hidden space representation. 
Its main idea is to update embedding $h_v$ for each node $v$ by aggregating its own features $h_v$ and its neighbors’ features $h_u$, where $u \in N(v)$, as Eq.~(\ref{equ: GCN-update}). 

\begin{equation}
\begin{aligned}
    h_v = f(\Theta^T \sum_{u \in \{ N(v) \cup  v\}} \bar{A}_{u,v}h_u),
\end{aligned}
\label{equ: GCN-update}
\end{equation}
where $f(\cdot)$ is an activation function, $\Theta$ is learned parameter, and $\bar{A}_{u,v}$ is correlation coefficient between node $u$ and node $v$ computed by the adjacency matrix \cite{kipf2016semi}.
In practice, we apply a 2-layered GCN to get the final schema embedding, thus one column embedding is related to its neighbor nodes (table) and nodes in the common table. The final schema embedding $Z_s \in \mathbb{R}^{l_{s} \times d_{model}}$ is obtained by
\begin{equation}
    Z_s = \texttt{GCN}(\texttt{GCN}(H_s)).
\label{equ:GCN-multi}
\end{equation}

\subsection{Speech-Schema Relation-aware Encoder}
\label{sec:sub-speech-schema-relation}

In traditional text-to-SQL tasks, schema-linking, which identifies references of tables and columns in the NL questions, could greatly improve the accuracy of the models \cite{lei2020re}. 
While it is trivial to identify the mentioned database keywords in an NL query, it's quite hard to assign the linking information between the speech and database in advance. 
As such, we design an advanced encoder to predict the relation information between speech and database automatically.

\subsubsection{Self-learned Schema Linking}
The goal of schema-linking is to recognize the items (table, column, and database values) mentioned in the speech question and then enhance the speech embedding by incorporating the embeddings from these correlated items. 
This module takes the speech embedding $Z_a$ and the schema embedding $Z_s$ as input and then outputs an advanced hybrid speech embedding. 
Concretely, a feed-forward network is applied to the speech and schema firstly, which aims to map them into a common hidden space. 
To learn a speech representation that considers its related schema items, we further perform an attention mechanism \cite{vaswani2017attention} over the speech frames and the schema items.
A semantic similarity score, denoted as $g_{a,s} \in \mathbb{R}^{l_{a} \times l_{s}}$ ($l_{a}$ is short for $l_{a}^{(N_a)}$, which denotes the length of speech in the output layer), and $g_{a,s}^{i,j}$ is calculated between $i$-th speech frame and $j$-th schema nodes to serve as a schema-linking relevance score, and then the weighted sum of the schema embedding is concatenated to the original speech embedding. Mathematically, the whole process is represented as follows.

\begin{equation}
\begin{aligned}
    g_{a,s}&=\frac{Z_a(Z_s)^{T}}{|| Z_a|| \cdot ||Z_s ||},
\end{aligned}
\end{equation}
\begin{equation}
\begin{aligned}
    C_a&=g_{a,s}Z_s,
\end{aligned}
\end{equation}
\begin{equation}
\begin{aligned}
    Z_{a} &= Z_a + C_a,
\end{aligned}
\end{equation}
where $C_a \in \mathbb{R}^{l_{a} \times d_{model}}$.

\subsubsection{Multi-modal Transformer-based Fusion}
To leverage both multi-modal and long-term temporal relationships among different modalities (i.e., audio and text), we learn a joint representation for the speech and the schema by a multi-modal transformer-based encoder. Inspired by the vanilla transformer structure \cite{vaswani2017attention}, the encoder is composed of a stack of basic blocks, each of which consists of a multi-head attention mechanism, a fully connected feed-forward network, and a layer normalization \cite{ba2016layer}. 
However, different from the basic transformer structure, we add two attention modules in our multi-modal transformer, the first of which is a self-attention (SA) module designed for learning the relationships inherent in one modality, while the latter of which is a crossmodal attention (CA), which is proposed to fuse the latent relation information among different modalities. 

Due to space limitations, we only illustrate the generation process of the improved speech embedding here, and the schema embedding is constructed similarly.
Given the inputs from the two modalities (i.e., the speech embedding $Z_a$ and the schema embedding $Z_s$), we define the speech query as $Q_a=Z_{a}W_{Q_a} \in \mathbb{R}^{l_{a} \times d_{model}}$, speech key as $K_a=Z_{a}W_{K_a} \in \mathbb{R}^{l_{a} \times d_{model}}$, speech value as $V_a=Z_{a}W_{V_a} \in \mathbb{R}^{l_{a} \times d_{model}}$, schema key as $K_s=Z_{s}W_{K_s} \in \mathbb{R}^{l_{a} \times d_{model}}$, and schema value as $V_s=Z_{s}W_{V_s} \in \mathbb{R}^{l_{a} \times d_{model}}$.

We first employ the SA module over the speech embedding to obtain $Y_a^{SA} \in \mathbb{R}^{l_{a} \times d_{model}}$, then perform the crossmodal attention between the speech and the schema to get $Y_a^{CA} \in \mathbb{R}^{l_{a} \times d_{model}}$. 
Furthermore, the sum of $Y_a^{SA}$ and $Y_a^{CA}$ is used to represent the multi-modal speech output $Y_a$, whose $i$-th time step is a weighted sum of $V_a$ and $V_s$, with the weight determined by a $\texttt{Softmax}(\cdot)$ score matrix. 
Finally, layer normalization and feed-forward network, which has been proved effective in improving the performance in prior works \cite{ba2016layer}, are also incorporated into the encoder. After the above-mentioned steps, we could get a new embedding for the speech and the schema, denoted as $Z_a$ and $Z_s$, respectively. We repeat this process $N_e$ times to obtain the final embedding. Mathematically, the whole process can be represented as:
\begin{equation}
\begin{aligned}
    Y_a^{SA} &= \texttt{MultiHead}(Q_a, K_a, V_a),
\end{aligned}
\end{equation}
\begin{equation}
\begin{aligned}
    Y_s^{CA} &= \texttt{MultiHead}(Q_a, K_a, V_s),
\end{aligned}
\end{equation}
\begin{equation}
\begin{aligned}
    Y_a &= Y_a^{\texttt{SA}} + Y_a^{\texttt{CA}},
\end{aligned}
\end{equation}
\begin{equation}
\begin{aligned}
    Y_a &= \texttt{LN}(Y_a + Z_a),
\end{aligned}
\end{equation}
\begin{equation}
\begin{aligned}
    Z_a &= \texttt{LN}(\texttt{FFN}(Y_a) + Y_a),
\end{aligned}
\end{equation}
where $\texttt{MultiHead}(\cdot)$ means the multi-head mechanism mentioned in section \ref{subsec:transformer}, $\texttt{LN}(\cdot)$ is the layer normalization, and $\texttt{FFN}(\cdot)$ refers to feed-forward network. 

\begin{figure}[t!]
    \centering
    \includegraphics[width=0.50\textwidth]{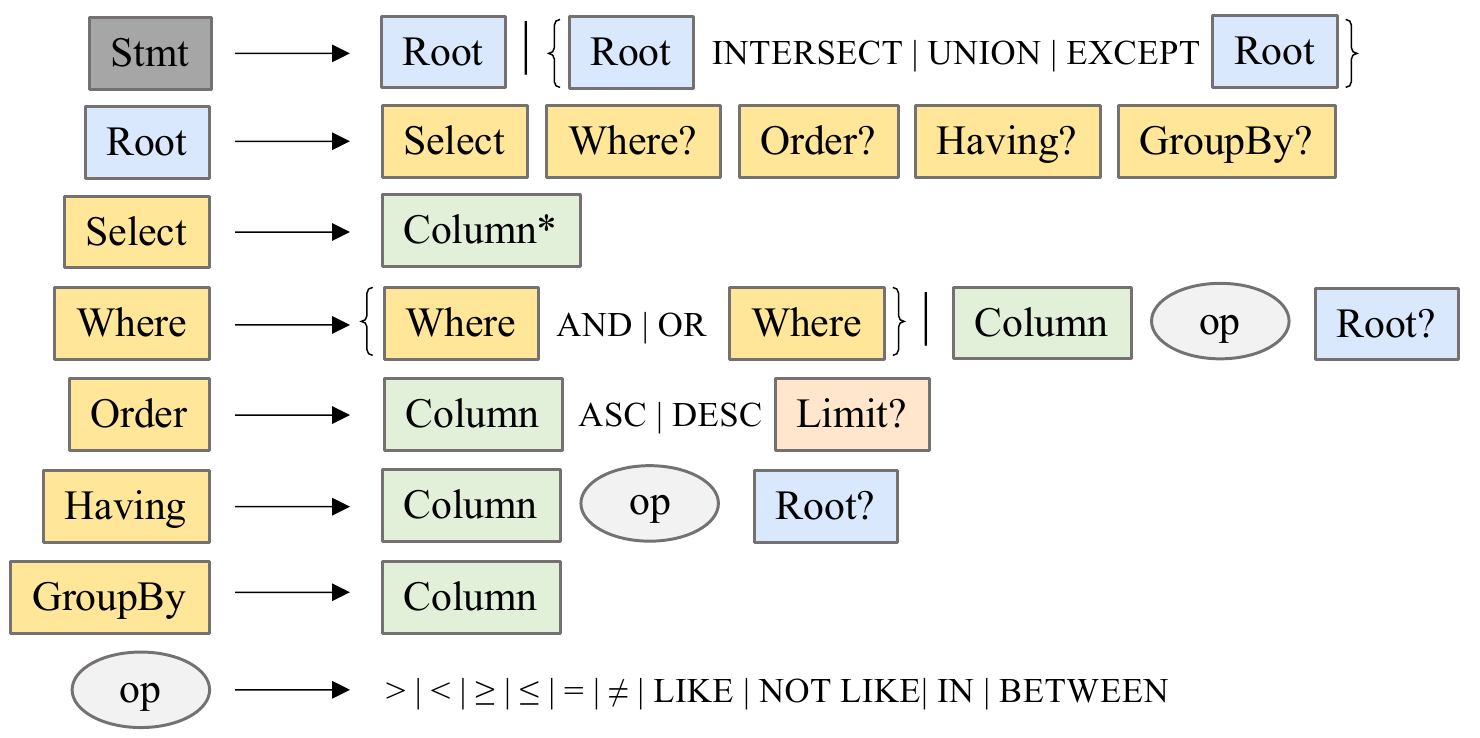}
    \caption{The Grammar of SemQL}
    \label{fig:grammer}
\vspace{-10pt}
\end{figure}

\subsection{SQL-aware Decoder}
\label{sec:sub-sql-decoder}

Since SQL is a programming language with clear and strict grammar, it would be better to encode this grammar information as a prior to guide the generation process. 
Following the practice in text-to-SQL \cite{guo2019towards}, in our implementation, we also choose the grammar of SemQL, as shown in Figure~\ref{fig:grammer}.
Our decoder is adapted from a grammar-based structure commonly-used in the text-to-SQL task \cite{guo2019towards}, which employs a LSTM architecture to synthesize SemQL by choosing a sequence of actions denoted as $\tilde{y} $. Mathematically, the generation process of a SemQL query sequence $\tilde{y} $ can be formalized as:
\begin{equation}
    p(\tilde{y} |x, S) = \prod_{i=1}^{T}p(a_i|x,S,a_{<i}),
\label{equ:ast-gen}
\end{equation}
where $a_i$ means an action applied at step $i$, $a_{<i}$ are all the previous actions before step $i$, and $T$ is the number of all the actions to predict $y$.
The actions involved in the generation process of Eq.~(\ref{equ:ast-gen}) is further categorized into two types: (i) \emph{ApplyRule}: applying a production rule to the current grammar tree till finishing the SQL sketch; (ii) \emph{SelectSchema}: selecting a column or table item from the schema to complete the SQL query. 

\subsubsection{ApplyRule} The objective of the \emph{ApplyRule} step is to construct a context-free grammar tree of the SQL query \cite{guo2019towards}. 
In every step, we select the most probable branch given the previous route, and employ an LSTM-based structure to simulate the process. 
At each prediction step $i$, the LSTM state is updated based on previous state $h_{i-1} \in \mathbb{R}^{d_{model}}$, previous action embedding $a_{i-1} \in \mathbb{R}^{d_{action}}$ ($d_{action}$ is the size of the action embedding), previous action type embedding $n_{i-1} \in \mathbb{R}^{d_{type}}$ ($d_{type}$ is the size of the action type embedding), and previous context representation of LSTM $c_{i-1}$. 
Then a Luong attention mechanism \cite{luong2015effective} is implemented where all of the speech final embeddings $Z_a$ would be taken into computation, finally the probability of selecting a rule is calculated by Eq.~(\ref{equ:decoder-sketch}).
\begin{equation}
\begin{aligned}
    h_i &= \texttt{LSTM}([a_{i-1};n_{i-1};c_{i-1}], h_{i-1}),
\end{aligned}
\end{equation}
\begin{equation}
\begin{aligned}
    c_i &= \texttt{Softmax}(h^T_{i}W_{a}Z^T_a )Z_a,
\end{aligned}
\end{equation}
\begin{equation}
\begin{aligned}
    u_i &= \texttt{tanh}([h_i; c_{i}]W_{u} + b_u),
\end{aligned}
\end{equation}
\begin{equation}
\begin{aligned}
    p(\tilde{y}_i=a_i | x, S, a_{<i}) &= \texttt{Softmax}(\texttt{tanh}(u^T_{i}W_{p} + b_p)),
\end{aligned}    
\label{equ:decoder-sketch}
\end{equation}
where $[;]$ refers to concatenation operation, $W_a \in \mathbb{R}^{d_{model} \times d_{model}}$, $W_u \in \mathbb{R}^{2d_{model} \times {d_{model}}}$ are the trainable weights, $W_p \in \mathbb{R}^{d_{model} \times n_{a}}$ ($n_{a}$ is the number of actions and $n_{a}$=46 in our setting),
$b_c$ is trainable bias, and the initial state $h_0$ is obtained by a max-pooling operation of the speech embedding $Z_a$.

\subsubsection{SelectSchema}
To fill in the specific item in the target SQL, another LSTM-based structure is also employed. 
In this part, we need to predict the operation (e.g., \texttt{max}, \texttt{min}, \texttt{count}) and the item (column or table) involved in the speech given the schema. 
The schema varies in every case and the desired items are also not fixed, which is quite different from the \emph{ApplyRule} part.
As such, we further employ the pointer network \cite{vinyals2015pointer} to handle this issue. 
The probability of selecting a schema item is calculated as follows 
\begin{equation}
\begin{aligned}
    p(\tilde{y}_i=a_i | x, s, a_{<i}) &= \texttt{Softmax}(u^{T}_{i}W_sZ^T_s),
\end{aligned}    
\label{equ:decoder}
\end{equation}
where $W_{s} \in \mathbb{R}^{d_{model} \times d_{model}}$ is a trainable weight matrix.
In particular, table options are restricted to the tables where the selected column is corresponding to.

\section{Model Training}
\label{subsec:pre-training}

Even though we have designed several modules to fuse the information from different modalities, there still exists a large gap between the speech modality and the text modality. 
To further minimize this modality gap, we propose a two-step framework (i.e., pre-training and fine-tuning) to train this network. The network first conducts model pre-training (Section~\ref{subsec:schema-linking} and Section~\ref{subsec:speech-item}) to align a common hidden space for each pair of speech and its corresponding transcript sentence, and then conducts fine-tuning (Section~\ref{subsec:fine-tune}) based on the datasets of speech-SQL pairs. 
It should be noticed that the pre-training step is optional, depending on the availability of the transcripts. 
Furthermore, the dataset used for the pre-training step is not necessarily required in the SQL domain.

\subsection{Speech-Sentence Pre-Training}
\label{subsec:schema-linking}
\begin{figure}[t!]
    \centering
    \includegraphics[width=0.50\textwidth]{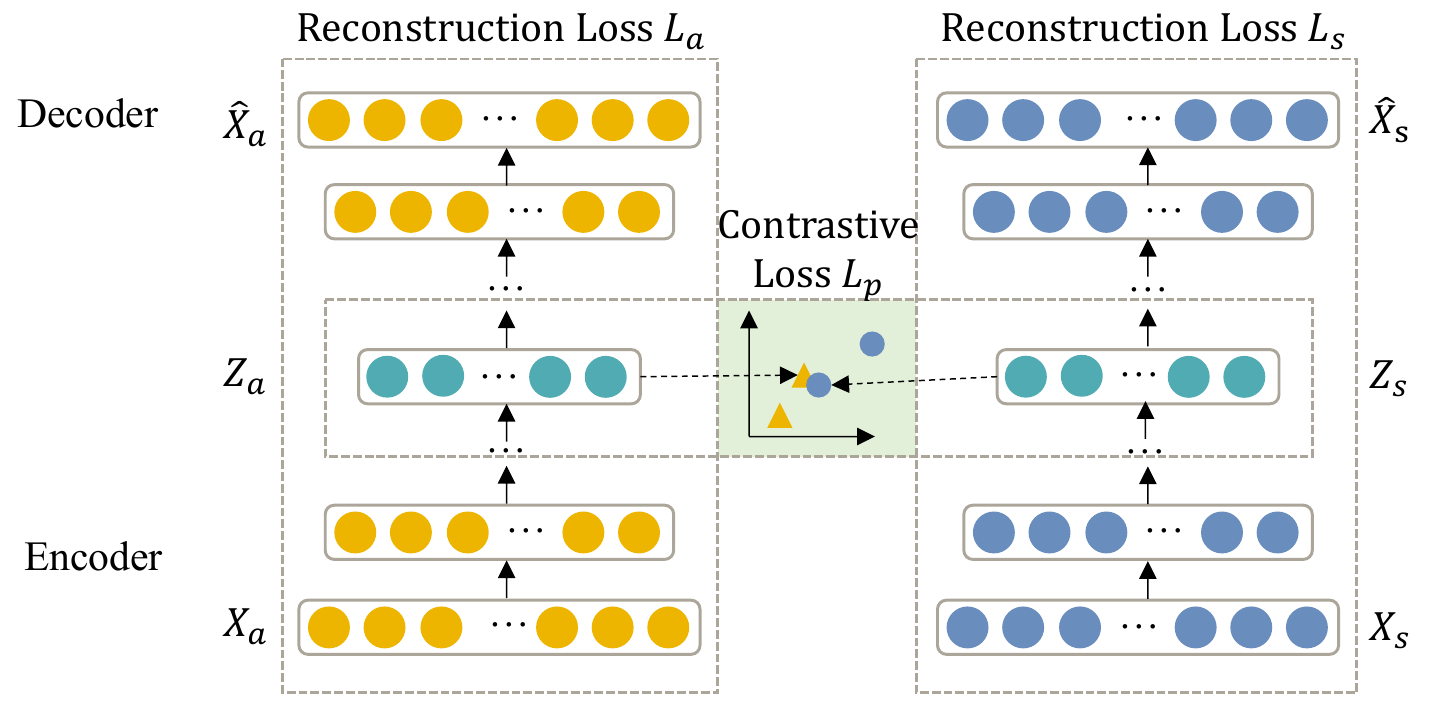}
    \caption{The Network Structure in the Speech-Sentence Pre-training Step.}
    \label{fig:pretrain-model}
\vspace{-10pt}
\end{figure}

As shown in Figure~\ref{fig:pretrain-model}, we design an autoencoder (AE)-based framework to match speech and text sentence with two different AEs. 
Semantic representations for different modalities are mapped by a speech autoencoder (SAE) and a text autoencoder (TAE) respectively, and these two representations are forced to be close to each other. 
Since we employ the same encoder like the one mentioned in Section~\ref{sec:sub-speech-encoder} for the SAE, and the same encoder as the one mentioned in Section~\ref{sec:sub-schema-encoder} for the TAE, the network parameters after this pre-training step can be reused in the following fine-tuning phase. 
And the decoders of the SAE and the TAE have reversed modules of their encoders according to the design rules of AEs. 
The loss of the network $\mathcal{L}$ is composed of three parts: a reconstruction loss of speech input $\mathcal{L}_a$, a reconstruction loss of transcript input $\mathcal{L}_s$, and a contrastive loss between the speech and text pairs $\mathcal{L}_p$, which can be formulated as follows.
\begin{equation}
\begin{aligned}
\mathcal{L}(X_a, \tilde{X}_s) = \mathcal{L}_a + \mathcal{L}_s + \mathcal{L}_p ,
\end{aligned}
\end{equation}
\begin{equation}
\begin{aligned}
    \mathcal{L}_a &=D_{KL}(X_a || \hat{X} _a),
\end{aligned}
\end{equation}
\begin{equation}
\begin{aligned}
    \mathcal{L}_s &=D_{KL}(\tilde{X}_s || \hat{X} _s),
\end{aligned}
\end{equation}
\begin{equation}
\begin{aligned}
    \mathcal{L}_p &=-\log \frac{exp(sim(h_a^b,h_s^b))}{\sum_{i[i\ne b]} \exp(sim(h_a^b, h_s^i))}, \\ 
\end{aligned}    
\label{equ:loss-pretrain}
\end{equation}
\begin{equation}
\begin{aligned}
sim(Z_a,Z_s) & =\frac{h_a^Th_s}{\left \| h_a \right \| \cdot \left \|  h_s\right \| },
\end{aligned}
\end{equation}
where $X_a$ and $\tilde{X}_s$ are the inputs of speech and transcript, $\hat{X}_a$ and $\hat{X}_a$ are the reconstructed outputs by the AEs, $D_{KL}(\cdot||\cdot)$ is a KL divergence used to measure the reconstruction loss. 
For contrastive loss, suppose a minibatch $G=\left \{(X_a^b, \tilde{X}_s^b)\right \} _{b=1}^{N_b}$ with $N_b$ examples is training simultaneously, and $h_a^b$ and $h_s^b$ are the intermediate representation of an example $b$ where $h_a^b$ is produced by a max-pooling operation from $Z_a$. And the loss is designed to shorten the distance between the same text and speech pair, and enlarge the distance between different text and speech pair. 
After training the AEs, embeddings extracted from speech and schema encoder can be regarded as close enough to each other.

\subsection{Speech-Item Pre-training}
\label{subsec:speech-item}

The objective of the speech-item pre-training phase serves as the same objective of the schema-linking step in the text-to-SQL task, that is, to explicitly identify if an item from the database schema is refereed by the speech question. 
Given the speech query and a item from the schema, their embeddings are first extracted by the speech and text encoder trained in the above-mentioned pre-training phase. Then a cosine-similarity following by Sigmoid nonlinearity function is employed to predict the existence as Eq.~(\ref{equ:finetuning}). 
\begin{equation}
\begin{aligned}
 \mathbf{\hat{y_f}} = \texttt{Sigmoid}(W_{f}\texttt{sim}(Z_a,h_s)+b_f)),
\label{equ:finetuning}
\end{aligned}
\end{equation}
where $W_f \in \mathbb{R}^{l_{q}}$ and $b_f$ are trainable parameters.

The training dataset for this pre-training step can easily be obtained from the original speech-to-SQL dataset. 
For any instance $\{\mathbf{x}, \mathbf{y}, \mathbf{S}\}$ in the speech-to-SQL dataset, we could compose multiple new training instances, each of which can be denoted as $\{\mathbf{x}, c, y_f\}$, where $c$ is a column item from the schema $S$, the label $y_f$ equals to 1 when $c$ appears in SQL $y$, otherwise 0. 
Finally, we employ the cross-entropy loss to in this pre-training step, defined as
\begin{equation}
\begin{aligned}
  \mathcal{L}(\mathbf{y_f}, \mathbf{\hat{y_f}}) = - (\mathbf{y_f^{T}} \log \mathbf{\hat{y_f}} + \mathbf{(1-y_f^{T})} \log \mathbf{(1-\hat{y_f}})),
\end{aligned}
\label{equ:loss}
\end{equation}

\subsection{Model Fine-tuning}
\label{subsec:fine-tune}

After above-mentioned pre-training phases, we obtain trained weights for speech encoder and schema text encoder. During this fine-tuning step, we will first load the weights by the pre-training step in advance if they exists, and then train the model globally with speech-to-SQL data, maximizing the log-likelihood of ground true action sequences defined as:
\begin{equation}
\begin{aligned}
  \mathcal{L} &= \max \sum_{(x,S,y) \in \mathcal{D}} \sum_{a_i \in ApplyRule} \log p(\tilde{y}_i = a_i | x, S, a_{<i}) \\
   &+ \sum_{a_i \in SelectSchema} \log p(\tilde{y}_i = a_i | x, S, a_{<i})).
\end{aligned}
\label{equ:loss}
\end{equation}
The whole network is trained in an end-to-end style with stochastic gradient descent methods such as Adam \cite{kingma2014adam}. 

\section{Experiments}
\label{sec:exp}

In this section, we evaluate the performance of the proposed model in terms of quantitative metrics. 
We first describe the experimental setup, including baselines in Section~\ref{sub:baseline},  implementation details in Section~\ref{sub:details}, and evaluation measurements in Section~\ref{sub:metrics}. 
Then we present the results in Section~\ref{sub:results} to demonstrate the effectiveness of our proposed models by comparing them with these baselines. 

\subsection{Baselines}
\label{sub:baseline}

There is currently no literature that achieves direct SQL generation from arbitrary speech questions. Existing text-to-SQL approaches assume the availability of the transcripts for each speech, either labeled by human annotators or recognized by an ASR system. In our experiments, we compare carefully designed methods to this problem, which can be roughly categorized into the cascaded approach and the end-to-end approach. For the cascaded approach, the discussed well-trained ASR system is first used to decode the transcripts for each speech, and then a text-to-SQL model is adopted to generate the SQL queries. Specifically, we adopt the following four cascaded baselines. 
\begin{itemize}
    \item \underline{ASR + Seq2SQL}: Seq2SQL \cite{zhong2017seq2sql} is the first DNN-based approach to solve text-to-SQL problem. The model is a straightforward Seq2Seq neural network, which takes the text as the input and the SQL statement as the output. 
    \item \underline{ASR + SQLNet}: SQLNet \cite{xu2018sqlnet} employs a more refined slot filling strategy. It takes advantage of the fact that most of the queries
in the WikiSQL dataset can be represented in a standard ``\texttt{\textup{SELECT \_ FROM \_ WHERE \_}}'' format.
    \item \underline{ASR + IRNet}: IRNet \cite{guo2019towards} is an advanced text-to-SQL model that improves the above two methods by representing the SQL statement with the SemSQL grammar in an abstract syntax tree (AST) format. This representation captures the structure information of the SQL generation problem and achieves much better results than the previous methods. 
    \item \underline{ASR + EditSQL}: EditSQL \cite{zhang2019editing} is another advanced text-to-SQL model that performs very well on SQL generation task. 
    This model could reuse previously generation results at the token level, which makes it a very strong baseline for performance comparison.
\end{itemize}
For the end-to-end approach, we design and implement the following three baselines. 
\begin{itemize}
    \item \underline{SpeechSeq2SQL}: This baseline employs a vanilla Seq2Seq model directly trained on the speech-to-SQL dataset that converts the speech signals into its corresponding SQL. The encoder and the decoder are both Bi-LSTM, and we name this method SpeechSeq2SQL.
    \item \underline{Transformer}: Transformer has shown quite promising performance on various NLP tasks from machine translation \cite{lakew2018comparison,currey2019incorporating}, dialogue system \cite{zhao2020multiple,le2019choi}, to ASR \cite{zeyer2019comparison,zhou2018comparison}. 
    In our experiment, we also implement the transformer structure according to \cite{vaswani2017attention}, with the speech question as the input and the SQL query as the output. 
    \item \underline{SpeechSQLNet}: this method is the end-to-end model that we designed in this work, which employs novel encoders, SQL-aware decoder, and pre-training mechanisms that are dedicated to this speech-to-SQL task. 
\end{itemize}

To ensure fairness and reproducibility, we train all the models with the same training set, tune their parameters with the same validation set, and finally evaluate the performance on the same testing set. 
We try our best to tune the parameters of these models to achieve their best performance. 

\subsection{Implementation Details} 
\label{sub:details}

The speech encoder is composed of a 6-layer CNN module with 1 input channel and 128 output channels, which takes speech features of the same size by resampling and zero-padding as needed. 
The schema encoder takes word embedding as inputs, which is initialized with a BiLSTM with 512 hidden units, and the outputs are produced by a 2-layers GCN. 
The transformer has a 2-layer encoder, with head size, feed-forward size, and hidden size set to be 4, 1024, 512 respectively. 
The dimension of action embedding and type embedding in the SQL-aware decoder are all set to 12. 
In addition, the pre-training models keep the same size as the above settings. 
During the training period, the learning rate, decay rate, dropout rate, and batch size of the network are set to 0.0001, 0.8, 0.3, and 256 respectively.
All the experiments were conducted on a server with a 72 Intel Core Processor (Xeon), 314 GB memory, Tesla K80 GPU, and CentOS. 

\subsection{Evaluation Metrics}
\label{sub:metrics}

Two widely-used metrics in SQL generation task is used in our experiments to validate the effectiveness of the proposed model. 
\begin{itemize}
    \item \underline{Query-Match Accuracy}: This metric evaluates the percentage of matches between the generated SQL with the ground truth. 
    To alleviate the negative effect of condition order, we further decouple each SQL statement into several parts: ``select column'', ``aggregator'', and ``condition''. Instead of directly comparing the SQL string, we take the condition part as a set and compare each element in the set. 
    \item \underline{Average Time Per Query (TPQ)}: This metric evaluates the averaged inference time cost of a query by various methods, and it reflects how fast each model can process the speech query.
\end{itemize}

\subsection{Experimental Results}
\label{sub:results}

\subsubsection{Comparison of Accuracy}
\label{sec:accuracy}

\begin{table}[t!]
\centering
\caption{Performance of Different Speech-to-SQL Methods.} 
   \begin{tabular}{l|cc|cc}
    \toprule

     \multirow{2}{*}{\textbf{Method}} & \multicolumn{2}{c}{\textbf{Validation}} & \multicolumn{2}{c}{\textbf{Test}} \\ [0.5ex]  
     &\textbf{Query Acc.}& \textbf{TPQ} &\textbf{Query Acc.}& \textbf{TPQ} \\ [0.5ex] 
    \midrule
    ASR + Seq2SQL & 0.0518 & 0.093  & 0.0568 & 0.089   \\
    ASR + Transformer & 0.0955 & 0.096  & 0.0782 & 0.094  \\
    ASR + IRNet  & 0.4264 & 0.146   & 0.4455 & 0.140 \\
    ASR + EditSQL  & 0.4973 & 0.128  & 0.5116 & 0.125 \\
    SpeechSeq2SQL  & 0.1918 & 0.003  & 0.1886 & 0.003  \\
    Transformer  & 0.1791 & 0.003  & 0.1805 & 0.004  \\
    SpeechSQLNet  & \textbf{0.5355} & 0.070  & \textbf{0.5415} & 0.070   \\
    \bottomrule
  \end{tabular}
  \label{tb:accuracy}
\end{table}

The accuracies of our proposed model together with the baseline on the dev and test datasets are presented in Table~\ref{tb:accuracy}, from which we could conclude the following observations.

The cascaded approaches do not perform very well and are not competitive. 
The main reason is that they suffer the error compounding problem between components, i.e., recognition errors leading to larger SQL conversion errors. 
Some end-to-end methods can outperform the cascaded methods by a large margin (e.g., ASR + Seq2SQL v.s SpeechSeq2SQL), proving the necessity of exploring approaches that bypass the text for the speech-to-SQL problem. 
The end-to-end approach has the advantage of naturally retaining the rich linguistic information in the speech, and conducting global optimization to reduce the errors. However, their capabilities to capture this linguistic information are not equally powerful. 
The performance of the baseline SpeechSeq2SQL method is less powerful than other end-to-end methods, and the reason is obvious. 
The SpeechSeq2SQL does not consider the specificity of the SQL generation problem, and hence has limited ability in both understanding the information converted in the speech and the database schema. 
The transformer-based structure could improve the performance compared with the basic SpeechSeq2SQL model, which is consistent with previous studies \cite{lakew2018comparison,zhao2020multiple}.
Among all these methods, the proposed method SpeechSQLNet uses a powerful speech encoder and a schema encoder to extract meanings from the speech and database schema, and hence it could beat all compared methods. Furthermore, the SQL-aware decoder can guide the generation process with the SQL grammar. 
This set of experiments verifies the rationale of the proposed speech-to-SQL problem and the effectiveness and necessity of the end-to-end approach. 

\subsubsection{Ablation Studies: The Effects of Each Network Component.}
\label{sec:Quantitative Comparison}

\begin{table}[t!]
\centering
\caption{Ablation Study Results.} 
\begin{tabular}{lcc}
    \toprule
    \textbf{Method} & \textbf{Validation Query Acc.} & \textbf{Test Query Acc.} \\ 
    \midrule
    SpeechSQLNet  & 0.5355 & 0.5415   \\
    { { } w/o GCN} & 0.5209 & 0.5164  \\
    { { } w/o Linking} & 0.5273 & 0.5195  \\
    { { } w/o Fusion} & 0.4600 & 0.4532  \\
    { { } w/o SQL} & 0.4472 & 0.4219  \\
    \bottomrule
  \end{tabular}
  \label{tb:ablation}
\vspace{-15pt}
\end{table}

In this section, we performed ablation studies to show the contribution of each component in SpeechSQLNet. 
Specifically, we first evaluate the SpeechSQLNet with all the components as the baseline. 
Each model is then represented by the name(s) of the components that it removes or replaces. 
To evaluate the effectiveness of the schema encoder, we replace it with a vanilla RNN-based encoder (\textbf{w/o GCN}). 
For the schema-linking component, we compared it with a baseline without it (\textbf{w/o Linking}). 
For the transformer-based fusion part, we remove it and name the baseline \textbf{w/o Fusion}.
Finally, for the SQL-aware decoder, we replace it with a basic RNN-based decoder (\textbf{w/o SQL}). 
All the results are listed in Table~\ref{tb:ablation}. 

Compared with the model without the schema-linking part, the performance decreases greatly, up to 10.76\% relative accuracy reduction. 
The significant reduction demonstrates the effectiveness of our proposed schema-linking mechanism in addressing the speech-to-SQL task. 
Other modules show similar conclusions. For example, the fusion mechanism brings around 16.40\% relative improvement, 
while the SQL-aware decoder shows around 28.35\% relative improvement.
However, compared with a vanilla RNN encoder, the GNN-based schema encoder only shows an 0.69\% relative improvement. 
The main reason is that the schema encoder mainly handles the complicated queres across tables, while most of the cases in our dataset focus on SQL queries on single tables (from WikiSQL). 
However, a further case study with multiple tables in the following Section~\ref{sec:case} shows the necessity of involving a GNN-based schema encoder in handling these complicated queries.

\subsubsection{Ablation Studies: The Effects of The Pre-training Mechanisms.}

We also conduct another set of ablation studies to show the effectiveness of the two proposed pre-training mechanisms.
Each model is still represented by the name(s) of the components that it removes, namely \textbf{w/o SSPT}, \textbf{w/o SIPT}, and \textbf{w/o Both}, referring to trained models without the \underline{s}peech-\underline{s}entence \underline{p}re-\underline{t}raining mechanism, without the \underline{s}peech-\underline{i}tem \underline{p}re-\underline{t}raining mechanism, and without these two pre-training mechanisms, respectively.  

As shown in Table~\ref{tb:ablation2}, these two pre-training mechanisms together bring around 15.29\% relative query match accuracy improvements in the test set, showing the necessity and effectiveness of involving these pre-training mechanisms to align the semantic representation for speech and text. 
Specifically, the speech-sentence pre-training mechanism solely brings about 4.97\% relative query match accuracy improvements in the test set, while the speech-item pre-training mechanism solely shows a 3.29\% relative query match accuracy improvement. 
Lastly, it should be noticed that, unlike the training step, the labeled data used in the pre-training step are not restricted to the SQL domain, and any labeled ASR dataset can be used. 

\begin{table}[t!]
\centering
\caption{The Effects of The Pre-training Mechanisms.} 
\begin{tabular}{lcc}
    \toprule
    \textbf{Method} & \textbf{Validation Query Acc.} & \textbf{Test Query Acc.} \\ 
    \midrule
    SpeechSQLNet  & 0.5355 & 0.5415    \\
    { { } w/o SSPT} & 0.5136 & 0.5027  \\
    { { } w/o SIPT} & 0.4982 & 0.5109 \\
     { { } w/o Both} & 0.4655 & 0.4577 \\
    \bottomrule
  \end{tabular}
  \label{tb:ablation2}
\vspace{-15pt}
\end{table}

\subsubsection{Hyper-parameter Study}
In this section, we study the performance variation affected by the parameters such as the number of GCN layers, head numbers, and the number of transformer layers. And the results are show in Figure~\ref{fig:hyper-parameter1}.

\begin{figure*}[th!]
  \centering
   \subfigure[{GCN Layer}]{
   \centering
     \includegraphics[width=0.31\textwidth]{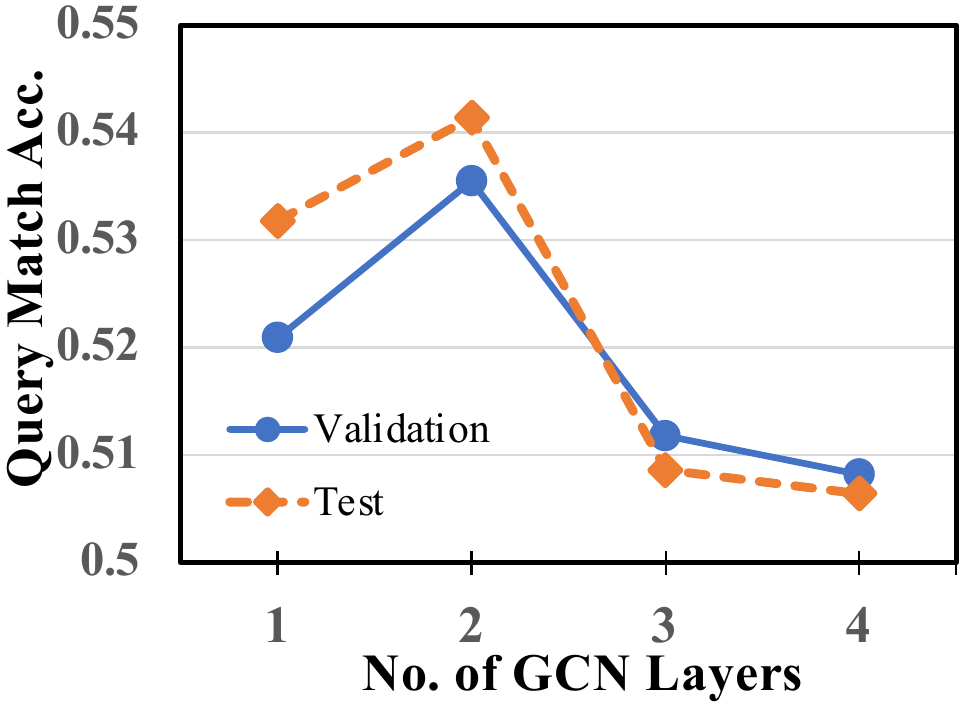}
     \label{fig:gcn}
   }
   \label{fig:hyper-parameter1}
  \subfigure[{Transformer Layer}]{
   \centering
     \includegraphics[width=0.31\textwidth]{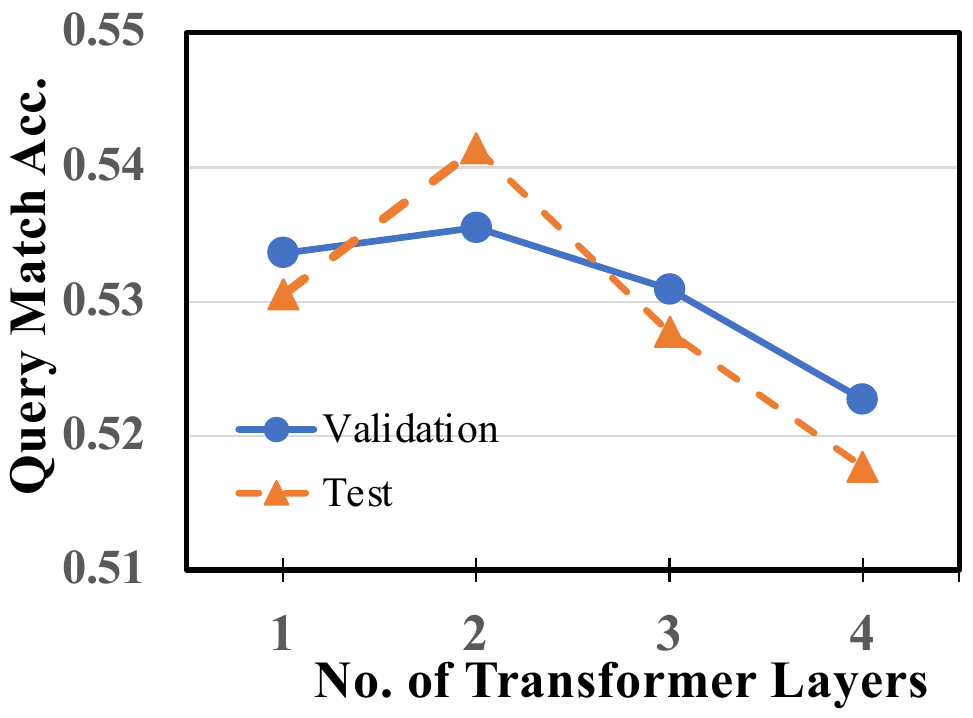}
     \label{fig:head}
  }
  \subfigure[{Transformer Head}]{
   \centering
     \includegraphics[width=0.31\textwidth]{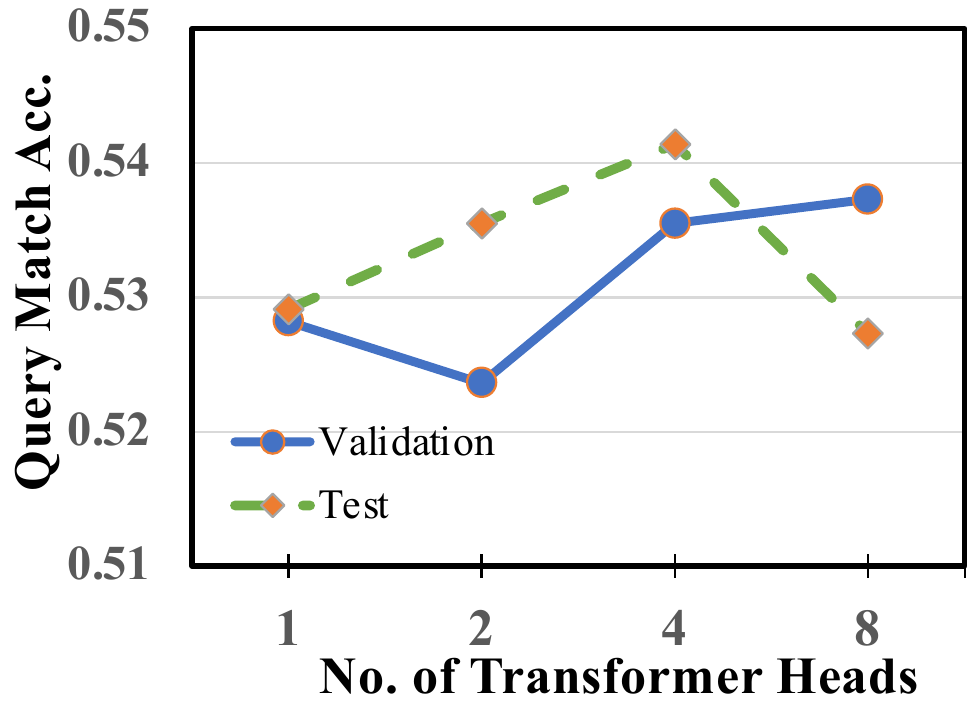}
     \label{fig:transformer}
  }
\vspace{-5pt}
  \caption{The Hyper-parameter Study of the SpeechSQLNet Model}
\label{fig:hyper-parameter1}
\vspace{-5pt}
\end{figure*}

As shown in Figure~\ref{fig:gcn}, a large number of GCN layers does not always lead to better performance. 
The performance reaches a peak at layer 2 and then decreases when the number exceeds this value. 
Previous studies on GNN \cite{chen2020handling} have also shown similar observations. And the reason is obvious, that is, the learning capacity of the model increases with the increasing of the layers, however, too many layers require more data to train \cite{li2019deepgcns}. 
Then for the other two parameters - the head numbers and the number of transformer layers, we also observe similar phenomena in Figure~\ref{fig:head} and Figure~\ref{fig:transformer}, respectively.

\subsubsection{Case Study}
\label{sec:case}

\begin{table*}[th!]
\centering
\caption{Two SQL Examples Generated by Various Cascaded and The End-to-end Methods.}
\vspace{-5pt}
\scalebox{0.86}{
\begin{tabular}{>{\centering\arraybackslash}m{2.4cm}|m{8.3cm}|m{9.0cm}}
\toprule
\makecell[l]{Speech Query} & \makecell[l]{What is the lowest number of draws with more than 1 byes?} & \makecell[l]{What is the name of the product with the color description 'yellow'?} \\
\makecell[l]{Schema} & \makecell[l]{wimmra(Wimmera\_fl,wins,byes,losses,draws,against)} & \makecell[l]{Products(product\_id, color\_code, product\_name, ...), \\ Ref\_Colors(color\_code, color\_description)} \\
\makecell[l]{Ground Truth SQL} & \makecell[l]{SELECT MIN(draws) FROM Wimmera WHERE byes $>$ 1 } & \makecell[l]{SELECT T1.product\_name FROM Products AS T1 JOIN Ref\_Colors \\ AS T2 WHERE T2.color\_description = 1} \\
\midrule
\makecell[l]{ASR Result} & \makecell[l]{what is the lowest number of drawers with more than one bites} & \makecell[l]{what is the name of the product with the color description yellow} \\
\midrule
\makecell[l]{ASR + Seq2SQL} & \makecell[l]{SELECT MIN(runners) FROM Boston WHERE draws $>$ 1} & \makecell[l]{SELECT name FROM Kazakhstan WHERE upstream = 1} \\ 
\midrule
\makecell[l]{ASR + Transformer} & \makecell[l]{SELECT MIN(sample\_size) FROM Martin WHERE \\ vote\_to\_evict = 1} & \makecell[l]{SELECT name FROM 2008 WHERE 2008\_club = 1} \\ 
\midrule
\makecell[l]{ASR + IRNet} & \makecell[l]{SELECT MIN(*) FROM Wimmera WHERE draws $>$ 1}   & \makecell[l]{SELECT T1.product\_name FROM Products AS T1 JOIN Ref\_Colors \\ AS T2 WHERE T2.color\_description = 1} \\ 
\midrule
\makecell[l]{ASR + EditSQL} & \makecell[l]{SELECT MIN(against) FROM Wimmera WHERE wins $>$ 1} & \makecell[l]{SELECT product\_name FROM Products WHERE color\_description = 1} \\ 
\midrule
\makecell[l]{SpeechSeq2SQL} & \makecell[l]{SELECT MIN(draws) FROM 1908 WHERE byes $>$ {1}} & \makecell[l]{SELECT name FROM Match\_season WHERE college =1} \\ 
\midrule
\makecell[l]{Transformer} & \makecell[l]{SELECT MIN(draws) FROM Golden WHERE byes $>$ 1} & \makecell[l]{SELECT name FROM List WHERE 2012\_club = \{value\}} \\ 
\midrule
\makecell[l]{SpeechSQLNet} & \makecell[l]{SELECT MIN(draws) FROM Wimmera WHERE byes $>$ 1} & \makecell[l]{SELECT T1.product\_name FROM Products AS T1 JOIN Ref\_Colors \\ AS T2 WHERE T2.color\_description = 1} \\ 
\bottomrule
\end{tabular}
}
\vspace{-5pt}
\label{fig:case}
\end{table*}

We list two cases to vividly show the generated SQLs by the various baselines as well as our model in Table~\ref{fig:case}. 
The first case illustrates the error pounding problem between the ASR and the text-to-SQL components. 
The cascaded methods fail since the ASR model misrecognized ``\textit{draws}'' to ``\textit{drawers}'' and ``\textit{byes}'' to ``\textit{bites}'', 
and thus the downstream text-to-SQL models all failed. 
This reflects the fact that the robustness of the cascaded solution is weak and the error compounding problem greatly affects the final SQL conversion accuracy.
However, the results generated by the end-to-end baselines such as SpeechSeq2SQL and Transformer are partly correct. 
Compared with all these baselines, our designed SpeechSQLNet could accurately capture the semantics conveyed by the speech question and then generates the SQL query accurately.

The second case shows how the schema could affect the predicted SQL. 
Since the required columns are from two tables, the desired SQL has a complicated `Join' operations between the `Product' and the `Ref\_Colors' tables. 
In this case, even the ASR module correctly recognizes the transcript, most of the cascaded approaches such as EditSQL and Transformer still fail to give the correct predictions. 
In contrast, the SpeechSQLNet could accurately identify the corresponding tables and columns, which validates the necessity of exploring end-to-end solutions for the Speech-to-SQL problem.

\subsubsection{Summary}
\label{sec:summary}

The experimental evaluations on the SpeechQL dataset demonstrate the superiority of SpeechSQLNet model over several strong baselines, including various cascaded ones as well as the end-to-end ones. 
For example, SpeechSQLNet achieves up to 11.96\% exact match accuracy improvements compared with the advanced IRNet model.
The effectiveness of each designed component is further validated by the ablation study. 
Specifically, the novel pre-training mechanisms bring around 15.29\% relative query match accuracy improvements in the test set.
The two generated examples in the case study section also vividly reflect the value of this problem and the superiority of our proposed method.  

\vspace{-2pt}
\section{Related Work}
\label{sec:related}

Our work is closely related to the research fields of text-to-SQL, speech-driven querying systems, and speech-to-X. 
We briefly survey the most related work from these three aspects.
\vspace{-5pt}
\subsection{Text-to-SQL}
Text-to-SQL aims to provide databases with a text-based interface. 
Past literature in this field can be classified into three categories: rule and template-based approach, statistical learning-based approach, and end-to-end neural network-based approach. Typical studies in rule and template-based approaches include works such as SODA \cite{blunschi2011data}, QUICK \cite{zenz2009keywords}, SINA \cite{shekarpour2015sina}, and NLQ/A \cite{zheng2017natural}. These approaches normally work in a similar style and suffer the drawback of poor flexibility since users may express their questions using different linguistic styles. For example, SODA \cite{blunschi2011data} adopts a five-step pipeline to translate a keyword-based NL input question into an SQL query. NLQ/A \cite{zheng2017natural} extends the input to pattern-based NL and supports more complex questions like concepts. 
A comprehensive survey on classic approaches can be found in \cite{affolter2019comparative}. The statistical learning-based approach can partially relieve the drawback and improve the flexibility, with typical studies in this category such as KBQA \cite{cui2019kbqa} and AMUSE \cite{hakimov2017amuse}. Recent trends in deep neural networks (DNNs) also promote the development on end-to-end neural network-based approach. Representative studies in this category include Seq2SQL \cite{zhong2017seq2sql}, SQLNet \cite{xu2017sqlnet}, TypeSQL \cite{yu2018typesql}, Syntax SQL \cite{sun2018semantic}, IRNET \cite{guo2019towards}, and its extensions such as NL2pSQL \cite{chen2019nl2psql}. There are also some public datasets released in the community to promote the research on this field, such as Spider \cite{yu2018spider}, TableQA \cite{sun2020tableqa}, and WikiSQL \cite{zhong2017seq2sql}. The Spider dataset is a small-scale cross-domain dataset that contains complex nested and join queries. 
WikiSQL is much larger in size, however, it mainly focuses on simple queries on limited domains. 
TableQA is another large-scale cross-domain NL2SQL dataset in the Chinese language.

While text-to-SQL can bridge the gap between SQL and NL and provide a text-based interface for DMBS, we believe that the voice-based interface is a much easier and faster way for interacting with DBMS. According to the user study in \cite{shah2020speakql}, the voice-based interface can enable users to compose SQL queries considerably faster by up to 6.7x compared to typing on a text-based interface in a tablet device. Hence, in this work, we take one step further to explore the possibility of a speech-driven interface for databases.

\subsection{Speech-driven Querying Systems}

Speech-driven querying systems have a long history with a substantial amount of industrial applications. 
To name a few, Nuance’s Dragon NaturallySpeaking\footnote{\url{https://www.nuance.com/dragon}} supports various basic voice commands. 
Mainstream search engines such as Google, Baidu, and Bing all provide voice-based inputs to search the web. 
With the popularity of smart devices and mobile phones, AI-powered assistants such as Xiaoice, Siri, Alexa, and Google Home also provide users interaction with the system based on voice, e.g., querying daily weather and traffic, keeping track of airline flight, and so on.

Speech-driven SQL-based systems have also been studied in the research community. For example, EchoQuery \cite{lyons2016making} aims to translate user's voice input into SQL queries. SpeakQL \cite{shah2020speakql,shah2019demonstration} supports a subset of SQL statements and enables users to interact with the system with a speech-based interface. 
TalkSQL \cite{obaidotalksql} implements a similar function to SpeakQL, which also works in a three-step manner, first allowing a user to input some voice command, then translating the inputs into SQL queries, and finally delivering execution results to the user. CiceroDB-Zero \cite{trummer2020demonstrating} enables participants to explore large data sets via voice interfaces. 

However, none of these studies focuses directly on SQL generation from arbitrary common NL questions expressed in speech, and they usually work in a cascaded manner. 
Furthermore, some of them restrict the spoken queries to be an NL-based version of SQL or its variants with a limited subset of SQL grammar, and thus, they still require users to have a professional background in SQL. 

\subsection{Speech-to-X}
Speech-to-X refers to a wide range of speech-driven tasks including speech-to-text, speech-to-image, speech-to-code, speech-to-model, and so on. Among all these tasks, speech-to-text is the most common one with dozens of existing studies. 
When the text refers to the transcript corresponding to the speech, this problem is also well-known as the ASR problem. 
Currently, the most commonly used ASR model is the hybrid model, which typically consists of two components: an acoustic model (AM) and a language model (LM). 
The AM translates the speech signals features (e.g., MFCC \cite{foote1997content}) into the corresponding phonemic representation, while the LM calculates the probability of the decoded word sequences from the natural language perspective. 
With the dominating performance of DNN-based models in NLP tasks, the end-to-end ASR system also becomes quite popular in the research community with studies such as LAS \cite{chan2016listen}, RNN Transducer \cite{rao2017exploring}, Attention-based Models \cite{watanabe2017hybrid,bahdanau2016end}, and RNN Transducer with Attention \cite{tian2019self}. However, in terms of ASR applications in the industry, the hybrid one still dominates the market due to its relatively good performance. In our work, we also employ the hybrid ASR with the text-to-SQL as the baseline method.

The speech-to-text problem also covers diverse applications besides ASR. Another task worth mentioning is the speech-to-text translation \cite{bansal2017towards}, where the ``text'' here refers to the translated text in another language different from the one in the source speech. 
The speech-to-text translation is very helpful especially for low-resource scenarios where neither machine translation nor ASR is available. 
Speech-to-speech translation \cite{wahlster2013verbmobil} takes one step further to directly generate speech in the target language, and it is applied in NLP applications such as cross-lingual dialogue systems. Besides text, dozens of studies have also explored the possibility of converting other modalities such as image (i.e., speech-to-image) \cite{li2020direct,wang2020s2igan,wang2021generating}, code (i.e., speech-to-code or programming-by-voice) \cite{lee2019disability,desilets2006voicecode}, and software model (i.e., voice-driven modeling) \cite{black2019voice}.

\smallskip
Different to be the above-mentioned studies, speech-to-SQL is an urgent task that aims to explore the information conveyed by human speech and convert it into a SQL statement. It can be considered to be a special case of the general speech-to-code problem. 
All aforementioned speech-driven applications, including speech-to-SQL, directly generate the target due to three reasons: (i) avoiding the possible error compounding between components in the cascaded systems; (ii) further reducing the error by retaining the rich linguistic information in the speech; and (iii) 
unlocking the power of advanced NLP technologies to the languages that lack a commonly used written form \cite{wang2021generating}.
We believe that speech-to-SQL would not only benefit the database area for providing user-friendly interfaces of DBMS, but also promote research of programming-by-voice in software engineering and enlarge the speech-to-X family. 

\section{Conclusion}
\label{sec:con}

In this paper, we propose a novel paradigm speech-driven interface for the relational database, together with its corresponding task speech-to-SQL, aiming to directly convert human speech into SQL queries. Cascaded methods, as well as end-to-end models named SpeechSQLNet, are proposed to solve this problem. 
Extensive experimental evaluations on the constructed corpus show that SpeechSQLNet can generate high-quality SQL queries, outperforming several competitive baselines. 

The experimental results also validate the rationale of the Speech-to-SQL problem. In the next step, we would like to explore more end-to-end methods on Speech-to-SQL. In particular, our work shows that speech signals contain rich extra-linguistic information that is not available from text, and it would be interesting to explore a more detailed inner structure of speech signals when designing speech-to-SQL models. We wish Speech-to-SQL would become a popular task as text-to-SQL, and will also show inspire further works on designing novel speech-driven applications. 

\begin{acks}
We are grateful to anonymous reviewers for their constructive comments on this draft.
\end{acks}

\balance
\bibliographystyle{ACM-Reference-Format}
\bibliography{speech}


\begin{thebibliography}{80}


\ifx \showCODEN    \undefined \def \showCODEN     #1{\unskip}     \fi
\ifx \showDOI      \undefined \def \showDOI       #1{#1}\fi
\ifx \showISBNx    \undefined \def \showISBNx     #1{\unskip}     \fi
\ifx \showISBNxiii \undefined \def \showISBNxiii  #1{\unskip}     \fi
\ifx \showISSN     \undefined \def \showISSN      #1{\unskip}     \fi
\ifx \showLCCN     \undefined \def \showLCCN      #1{\unskip}     \fi
\ifx \shownote     \undefined \def \shownote      #1{#1}          \fi
\ifx \showarticletitle \undefined \def \showarticletitle #1{#1}   \fi
\ifx \showURL      \undefined \def \showURL       {\relax}        \fi
\providecommand\bibfield[2]{#2}
\providecommand\bibinfo[2]{#2}
\providecommand\natexlab[1]{#1}
\providecommand\showeprint[2][]{arXiv:#2}

\bibitem[\protect\citeauthoryear{Affolter, Stockinger, and Bernstein}{Affolter
  et~al\mbox{.}}{2019}]%
        {affolter2019comparative}
\bibfield{author}{\bibinfo{person}{Katrin Affolter}, \bibinfo{person}{Kurt
  Stockinger}, {and} \bibinfo{person}{Abraham Bernstein}.}
  \bibinfo{year}{2019}\natexlab{}.
\newblock \showarticletitle{A comparative survey of recent natural language
  interfaces for databases}.
\newblock \bibinfo{journal}{\emph{The VLDB Journal}} \bibinfo{volume}{28},
  \bibinfo{number}{5} (\bibinfo{year}{2019}), \bibinfo{pages}{793--819}.
\newblock


\bibitem[\protect\citeauthoryear{Alateeq, Roantree, and Gurrin}{Alateeq
  et~al\mbox{.}}{2020}]%
        {alateeq2020voxento}
\bibfield{author}{\bibinfo{person}{Ahmed Alateeq}, \bibinfo{person}{Mark
  Roantree}, {and} \bibinfo{person}{Cathal Gurrin}.}
  \bibinfo{year}{2020}\natexlab{}.
\newblock \showarticletitle{Voxento: A Prototype Voice-controlled Interactive
  Search Engine for Lifelogs}. In \bibinfo{booktitle}{\emph{Proceedings of the
  Third Annual Workshop on Lifelog Search Challenge}}. \bibinfo{pages}{77--81}.
\newblock


\bibitem[\protect\citeauthoryear{Ba, Kiros, and Hinton}{Ba
  et~al\mbox{.}}{2016}]%
        {ba2016layer}
\bibfield{author}{\bibinfo{person}{Jimmy~Lei Ba}, \bibinfo{person}{Jamie~Ryan
  Kiros}, {and} \bibinfo{person}{Geoffrey~E Hinton}.}
  \bibinfo{year}{2016}\natexlab{}.
\newblock \showarticletitle{Layer normalization}.
\newblock \bibinfo{journal}{\emph{arXiv preprint arXiv:1607.06450}}
  (\bibinfo{year}{2016}).
\newblock


\bibitem[\protect\citeauthoryear{Bahdanau, Chorowski, Serdyuk, Brakel, and
  Bengio}{Bahdanau et~al\mbox{.}}{2016}]%
        {bahdanau2016end}
\bibfield{author}{\bibinfo{person}{Dzmitry Bahdanau}, \bibinfo{person}{Jan
  Chorowski}, \bibinfo{person}{Dmitriy Serdyuk}, \bibinfo{person}{Philemon
  Brakel}, {and} \bibinfo{person}{Yoshua Bengio}.}
  \bibinfo{year}{2016}\natexlab{}.
\newblock \showarticletitle{End-to-end attention-based large vocabulary speech
  recognition}. In \bibinfo{booktitle}{\emph{2016 ICASSP}}. IEEE,
  \bibinfo{pages}{4945--4949}.
\newblock


\bibitem[\protect\citeauthoryear{Bansal, Kamper, Lopez, and Goldwater}{Bansal
  et~al\mbox{.}}{2017}]%
        {bansal2017towards}
\bibfield{author}{\bibinfo{person}{Sameer Bansal}, \bibinfo{person}{Herman
  Kamper}, \bibinfo{person}{Adam Lopez}, {and} \bibinfo{person}{Sharon
  Goldwater}.} \bibinfo{year}{2017}\natexlab{}.
\newblock \showarticletitle{Towards speech-to-text translation without speech
  recognition}. In \bibinfo{booktitle}{\emph{Proceedings of the 15th Conference
  of the European Chapter of the Association for Computational Linguistics:
  Volume 2, Short Papers}}. \bibinfo{pages}{474--479}.
\newblock


\bibitem[\protect\citeauthoryear{Black, Rapos, and Stephan}{Black
  et~al\mbox{.}}{2019}]%
        {black2019voice}
\bibfield{author}{\bibinfo{person}{Dana Black}, \bibinfo{person}{Eric~J Rapos},
  {and} \bibinfo{person}{Matthew Stephan}.} \bibinfo{year}{2019}\natexlab{}.
\newblock \showarticletitle{Voice-Driven Modeling: Software Modeling Using
  Automated Speech Recognition}. In \bibinfo{booktitle}{\emph{2019 ACM/IEEE
  22nd International Conference on Model Driven Engineering Languages and
  Systems Companion (MODELS-C)}}. IEEE, \bibinfo{pages}{252--258}.
\newblock


\bibitem[\protect\citeauthoryear{Blunschi, Jossen, Kossmann, Mori, and
  Stockinger}{Blunschi et~al\mbox{.}}{2011}]%
        {blunschi2011data}
\bibfield{author}{\bibinfo{person}{Lukas Blunschi}, \bibinfo{person}{Claudio
  Jossen}, \bibinfo{person}{Donald Kossmann}, \bibinfo{person}{Magdalini Mori},
  {and} \bibinfo{person}{Kurt Stockinger}.} \bibinfo{year}{2011}\natexlab{}.
\newblock \showarticletitle{Data-thirsty business analysts need SODA: search
  over data warehouse}. In \bibinfo{booktitle}{\emph{Proceedings of the 20th
  ACM international conference on Information and knowledge management}}.
  \bibinfo{pages}{2525--2528}.
\newblock


\bibitem[\protect\citeauthoryear{Bogin, Berant, and Gardner}{Bogin
  et~al\mbox{.}}{2019}]%
        {bogin2019representing}
\bibfield{author}{\bibinfo{person}{Ben Bogin}, \bibinfo{person}{Jonathan
  Berant}, {and} \bibinfo{person}{Matt Gardner}.}
  \bibinfo{year}{2019}\natexlab{}.
\newblock \showarticletitle{Representing Schema Structure with Graph Neural
  Networks for Text-to-SQL Parsing}. In \bibinfo{booktitle}{\emph{Proceedings
  of the 57th Annual Meeting of the Association for Computational
  Linguistics}}. \bibinfo{pages}{4560--4565}.
\newblock


\bibitem[\protect\citeauthoryear{Chan, Jaitly, Le, and Vinyals}{Chan
  et~al\mbox{.}}{2016}]%
        {chan2016listen}
\bibfield{author}{\bibinfo{person}{William Chan}, \bibinfo{person}{Navdeep
  Jaitly}, \bibinfo{person}{Quoc Le}, {and} \bibinfo{person}{Oriol Vinyals}.}
  \bibinfo{year}{2016}\natexlab{}.
\newblock \showarticletitle{Listen, attend and spell: A neural network for
  large vocabulary conversational speech recognition}. In
  \bibinfo{booktitle}{\emph{2016 IEEE International Conference on Acoustics,
  Speech and Signal Processing (ICASSP)}}. IEEE, \bibinfo{pages}{4960--4964}.
\newblock


\bibitem[\protect\citeauthoryear{Chazan, Hoory, Cohen, and Zibulski}{Chazan
  et~al\mbox{.}}{2000}]%
        {chazan2000speech}
\bibfield{author}{\bibinfo{person}{Dan Chazan}, \bibinfo{person}{Ron Hoory},
  \bibinfo{person}{Gilad Cohen}, {and} \bibinfo{person}{Meir Zibulski}.}
  \bibinfo{year}{2000}\natexlab{}.
\newblock \showarticletitle{Speech reconstruction from mel frequency cepstral
  coefficients and pitch frequency}. In \bibinfo{booktitle}{\emph{2000 IEEE
  International Conference on Acoustics, Speech, and Signal Processing.
  Proceedings (Cat. No. 00CH37100)}}, Vol.~\bibinfo{volume}{3}. IEEE,
  \bibinfo{pages}{1299--1302}.
\newblock


\bibitem[\protect\citeauthoryear{Chen, Hwang, Choo, Ha, and Kim}{Chen
  et~al\mbox{.}}{2019}]%
        {chen2019nl2psql}
\bibfield{author}{\bibinfo{person}{Fuxiang Chen}, \bibinfo{person}{Seung-won
  Hwang}, \bibinfo{person}{Jaegul Choo}, \bibinfo{person}{Jung-Woo Ha}, {and}
  \bibinfo{person}{Sunghun Kim}.} \bibinfo{year}{2019}\natexlab{}.
\newblock \showarticletitle{Nl2PSQL: Generating pseudo-SQL queries from
  under-specified natural language questions}. In
  \bibinfo{booktitle}{\emph{Proceedings EMNLP-IJCNLP}}.
  \bibinfo{pages}{2603--2613}.
\newblock


\bibitem[\protect\citeauthoryear{Chen and Wong}{Chen and Wong}{2020}]%
        {chen2020handling}
\bibfield{author}{\bibinfo{person}{Tianwen Chen} {and} \bibinfo{person}{Raymond
  Chi-Wing Wong}.} \bibinfo{year}{2020}\natexlab{}.
\newblock \showarticletitle{Handling information loss of graph neural networks
  for session-based recommendation}. In \bibinfo{booktitle}{\emph{Proceedings
  of the 26th ACM SIGKDD International Conference on Knowledge Discovery \&
  Data Mining}}. \bibinfo{pages}{1172--1180}.
\newblock


\bibitem[\protect\citeauthoryear{Cho, van Merrienboer, G{\"u}l{\c{c}}ehre,
  Bahdanau, Bougares, Schwenk, and Bengio}{Cho et~al\mbox{.}}{2014}]%
        {cho2014learning}
\bibfield{author}{\bibinfo{person}{Kyunghyun Cho}, \bibinfo{person}{Bart van
  Merrienboer}, \bibinfo{person}{{\c{C}}aglar G{\"u}l{\c{c}}ehre},
  \bibinfo{person}{Dzmitry Bahdanau}, \bibinfo{person}{Fethi Bougares},
  \bibinfo{person}{Holger Schwenk}, {and} \bibinfo{person}{Yoshua Bengio}.}
  \bibinfo{year}{2014}\natexlab{}.
\newblock \showarticletitle{Learning Phrase Representations using RNN
  Encoder-Decoder for Statistical Machine Translation}. In
  \bibinfo{booktitle}{\emph{EMNLP}}.
\newblock


\bibitem[\protect\citeauthoryear{Cui, Xiao, Wang, Song, Hwang, and Wang}{Cui
  et~al\mbox{.}}{2019}]%
        {cui2019kbqa}
\bibfield{author}{\bibinfo{person}{Wanyun Cui}, \bibinfo{person}{Yanghua Xiao},
  \bibinfo{person}{Haixun Wang}, \bibinfo{person}{Yangqiu Song},
  \bibinfo{person}{Seung-won Hwang}, {and} \bibinfo{person}{Wei Wang}.}
  \bibinfo{year}{2019}\natexlab{}.
\newblock \showarticletitle{KBQA: learning question answering over QA corpora
  and knowledge bases}.
\newblock \bibinfo{journal}{\emph{arXiv preprint arXiv:1903.02419}}
  (\bibinfo{year}{2019}).
\newblock


\bibitem[\protect\citeauthoryear{Currey and Heafield}{Currey and
  Heafield}{2019}]%
        {currey2019incorporating}
\bibfield{author}{\bibinfo{person}{Anna Currey} {and} \bibinfo{person}{Kenneth
  Heafield}.} \bibinfo{year}{2019}\natexlab{}.
\newblock \showarticletitle{Incorporating source syntax into transformer-based
  neural machine translation}. In \bibinfo{booktitle}{\emph{Proceedings of the
  Fourth Conference on Machine Translation (Volume 1: Research Papers)}}.
  \bibinfo{pages}{24--33}.
\newblock


\bibitem[\protect\citeauthoryear{D{\'e}silets, Fox, and Norton}{D{\'e}silets
  et~al\mbox{.}}{2006}]%
        {desilets2006voicecode}
\bibfield{author}{\bibinfo{person}{Alain D{\'e}silets},
  \bibinfo{person}{David~C Fox}, {and} \bibinfo{person}{Stuart Norton}.}
  \bibinfo{year}{2006}\natexlab{}.
\newblock \showarticletitle{Voicecode: An innovative speech interface for
  programming-by-voice}. In \bibinfo{booktitle}{\emph{CHI'06 extended abstracts
  on Human factors in computing systems}}. \bibinfo{pages}{239--242}.
\newblock


\bibitem[\protect\citeauthoryear{Devlin, Chang, Lee, and Toutanova}{Devlin
  et~al\mbox{.}}{2019}]%
        {devlin2019bert}
\bibfield{author}{\bibinfo{person}{Jacob Devlin}, \bibinfo{person}{Ming-Wei
  Chang}, \bibinfo{person}{Kenton Lee}, {and} \bibinfo{person}{Kristina
  Toutanova}.} \bibinfo{year}{2019}\natexlab{}.
\newblock \showarticletitle{BERT: Pre-training of Deep Bidirectional
  Transformers for Language Understanding}. In
  \bibinfo{booktitle}{\emph{Proceedings of NAACL}}.
\newblock


\bibitem[\protect\citeauthoryear{Fan, Qian, Xie, and Soong}{Fan
  et~al\mbox{.}}{2014}]%
        {fan2014tts}
\bibfield{author}{\bibinfo{person}{Yuchen Fan}, \bibinfo{person}{Yao Qian},
  \bibinfo{person}{Feng-Long Xie}, {and} \bibinfo{person}{Frank~K Soong}.}
  \bibinfo{year}{2014}\natexlab{}.
\newblock \showarticletitle{TTS synthesis with bidirectional LSTM based
  recurrent neural networks}. In \bibinfo{booktitle}{\emph{Fifteenth annual
  conference of the international speech communication association}}.
\newblock


\bibitem[\protect\citeauthoryear{Foote}{Foote}{1997}]%
        {foote1997content}
\bibfield{author}{\bibinfo{person}{Jonathan~T Foote}.}
  \bibinfo{year}{1997}\natexlab{}.
\newblock \showarticletitle{Content-based retrieval of music and audio}. In
  \bibinfo{booktitle}{\emph{Multimedia Storage and Archiving Systems II}},
  Vol.~\bibinfo{volume}{3229}. International Society for Optics and Photonics,
  \bibinfo{pages}{138--147}.
\newblock


\bibitem[\protect\citeauthoryear{Glorot, Bordes, and Bengio}{Glorot
  et~al\mbox{.}}{2011}]%
        {glorot2011deep}
\bibfield{author}{\bibinfo{person}{Xavier Glorot}, \bibinfo{person}{Antoine
  Bordes}, {and} \bibinfo{person}{Yoshua Bengio}.}
  \bibinfo{year}{2011}\natexlab{}.
\newblock \showarticletitle{Deep sparse rectifier neural networks}. In
  \bibinfo{booktitle}{\emph{Proceedings of the fourteenth international
  conference on artificial intelligence and statistics}}. JMLR Workshop and
  Conference Proceedings, \bibinfo{pages}{315--323}.
\newblock


\bibitem[\protect\citeauthoryear{Guo, Zhan, Gao, Xiao, Lou, Liu, and Zhang}{Guo
  et~al\mbox{.}}{2019}]%
        {guo2019towards}
\bibfield{author}{\bibinfo{person}{Jiaqi Guo}, \bibinfo{person}{Zecheng Zhan},
  \bibinfo{person}{Yan Gao}, \bibinfo{person}{Yan Xiao},
  \bibinfo{person}{Jian-Guang Lou}, \bibinfo{person}{Ting Liu}, {and}
  \bibinfo{person}{Dongmei Zhang}.} \bibinfo{year}{2019}\natexlab{}.
\newblock \showarticletitle{Towards Complex Text-to-SQL in Cross-Domain
  Database with Intermediate Representation}. In
  \bibinfo{booktitle}{\emph{Proceedings of the 57th Annual Meeting of the
  Association for Computational Linguistics}}. \bibinfo{pages}{4524--4535}.
\newblock


\bibitem[\protect\citeauthoryear{Hakimov, Jebbara, and Cimiano}{Hakimov
  et~al\mbox{.}}{2017}]%
        {hakimov2017amuse}
\bibfield{author}{\bibinfo{person}{Sherzod Hakimov}, \bibinfo{person}{Soufian
  Jebbara}, {and} \bibinfo{person}{Philipp Cimiano}.}
  \bibinfo{year}{2017}\natexlab{}.
\newblock \showarticletitle{AMUSE: multilingual semantic parsing for question
  answering over linked data}. In \bibinfo{booktitle}{\emph{International
  Semantic Web Conference}}. Springer, \bibinfo{pages}{329--346}.
\newblock


\bibitem[\protect\citeauthoryear{Ioffe and Szegedy}{Ioffe and Szegedy}{2015}]%
        {ioffe2015batch}
\bibfield{author}{\bibinfo{person}{Sergey Ioffe} {and}
  \bibinfo{person}{Christian Szegedy}.} \bibinfo{year}{2015}\natexlab{}.
\newblock \showarticletitle{Batch normalization: Accelerating deep network
  training by reducing internal covariate shift}. In
  \bibinfo{booktitle}{\emph{International conference on machine learning}}.
  PMLR, \bibinfo{pages}{448--456}.
\newblock


\bibitem[\protect\citeauthoryear{Kedar}{Kedar}{2009}]%
        {kedar2009database}
\bibfield{author}{\bibinfo{person}{Seema Kedar}.}
  \bibinfo{year}{2009}\natexlab{}.
\newblock \bibinfo{booktitle}{\emph{Database Management System}}.
\newblock \bibinfo{publisher}{Technical Publications}.
\newblock


\bibitem[\protect\citeauthoryear{Kim, So, Han, and Lee}{Kim
  et~al\mbox{.}}{2020}]%
        {kim2020natural}
\bibfield{author}{\bibinfo{person}{Hyeonji Kim}, \bibinfo{person}{Byeong-Hoon
  So}, \bibinfo{person}{Wook-Shin Han}, {and} \bibinfo{person}{Hongrae Lee}.}
  \bibinfo{year}{2020}\natexlab{}.
\newblock \showarticletitle{Natural language to SQL: Where are we today?}
\newblock \bibinfo{journal}{\emph{Proceedings of the VLDB Endowment}}
  \bibinfo{volume}{13}, \bibinfo{number}{10} (\bibinfo{year}{2020}),
  \bibinfo{pages}{1737--1750}.
\newblock


\bibitem[\protect\citeauthoryear{Kingma and Ba}{Kingma and Ba}{2014}]%
        {kingma2014adam}
\bibfield{author}{\bibinfo{person}{Diederik~P Kingma} {and}
  \bibinfo{person}{Jimmy Ba}.} \bibinfo{year}{2014}\natexlab{}.
\newblock \bibinfo{booktitle}{\emph{Adam: A Method for Stochastic
  Optimization}}.
\newblock \bibinfo{type}{{T}echnical {R}eport}.
\newblock


\bibitem[\protect\citeauthoryear{Kipf and Welling}{Kipf and Welling}{2016}]%
        {kipf2016semi}
\bibfield{author}{\bibinfo{person}{Thomas~N Kipf} {and} \bibinfo{person}{Max
  Welling}.} \bibinfo{year}{2016}\natexlab{}.
\newblock \showarticletitle{Semi-supervised classification with graph
  convolutional networks}.
\newblock \bibinfo{journal}{\emph{arXiv preprint arXiv:1609.02907}}
  (\bibinfo{year}{2016}).
\newblock


\bibitem[\protect\citeauthoryear{Krizhevsky, Sutskever, and Hinton}{Krizhevsky
  et~al\mbox{.}}{2012}]%
        {krizhevsky2012imagenet}
\bibfield{author}{\bibinfo{person}{Alex Krizhevsky}, \bibinfo{person}{Ilya
  Sutskever}, {and} \bibinfo{person}{Geoffrey~E Hinton}.}
  \bibinfo{year}{2012}\natexlab{}.
\newblock \showarticletitle{Imagenet classification with deep convolutional
  neural networks}.
\newblock \bibinfo{journal}{\emph{Advances in neural information processing
  systems}}  \bibinfo{volume}{25} (\bibinfo{year}{2012}),
  \bibinfo{pages}{1097--1105}.
\newblock


\bibitem[\protect\citeauthoryear{Kumar, Kumar, de~Boissiere, Gestin, Teoh,
  Sotelo, de~Brebisson, Bengio, and Courville}{Kumar et~al\mbox{.}}{2019}]%
        {kumar2019melgan}
\bibfield{author}{\bibinfo{person}{Kundan Kumar}, \bibinfo{person}{Rithesh
  Kumar}, \bibinfo{person}{Thibault de Boissiere}, \bibinfo{person}{Lucas
  Gestin}, \bibinfo{person}{Wei~Zhen Teoh}, \bibinfo{person}{Jose Sotelo},
  \bibinfo{person}{Alexandre de Brebisson}, \bibinfo{person}{Yoshua Bengio},
  {and} \bibinfo{person}{Aaron Courville}.} \bibinfo{year}{2019}\natexlab{}.
\newblock \showarticletitle{MelGAN: Generative Adversarial Networks for
  Conditional Waveform Synthesis}.
\newblock  (\bibinfo{year}{2019}).
\newblock


\bibitem[\protect\citeauthoryear{Lakew, Cettolo, and Federico}{Lakew
  et~al\mbox{.}}{2018}]%
        {lakew2018comparison}
\bibfield{author}{\bibinfo{person}{Surafel~Melaku Lakew},
  \bibinfo{person}{Mauro Cettolo}, {and} \bibinfo{person}{Marcello Federico}.}
  \bibinfo{year}{2018}\natexlab{}.
\newblock \showarticletitle{A Comparison of Transformer and Recurrent Neural
  Networks on Multilingual Neural Machine Translation}. In
  \bibinfo{booktitle}{\emph{Proceedings of the 27th International Conference on
  Computational Linguistics}}. \bibinfo{pages}{641--652}.
\newblock


\bibitem[\protect\citeauthoryear{{\L}a{\'n}cucki}{{\L}a{\'n}cucki}{2020}]%
        {lancucki2020fastpitch}
\bibfield{author}{\bibinfo{person}{Adrian {\L}a{\'n}cucki}.}
  \bibinfo{year}{2020}\natexlab{}.
\newblock \showarticletitle{Fastpitch: Parallel text-to-speech with pitch
  prediction}.
\newblock \bibinfo{journal}{\emph{arXiv preprint arXiv:2006.06873}}
  (\bibinfo{year}{2020}).
\newblock


\bibitem[\protect\citeauthoryear{LE, SAHOO, and CHEN}{LE
  et~al\mbox{.}}{[n.d.]}]%
        {le2019choi}
\bibfield{author}{\bibinfo{person}{Hung LE}, \bibinfo{person}{Doyen SAHOO},
  {and} \bibinfo{person}{Nancy~F CHEN}.} \bibinfo{year}{[n.d.]}\natexlab{}.
\newblock \showarticletitle{Multimodal transformer networks for end-to-end
  video-grounded dialogue systems.(2019)}. In
  \bibinfo{booktitle}{\emph{Proceedings of the 57th Annual Meeting of the
  Association for Computational Linguistics, Florence, Italy, 2019 July
  28-August}}, Vol.~\bibinfo{volume}{2}. \bibinfo{pages}{5612--5623}.
\newblock


\bibitem[\protect\citeauthoryear{Lee, Fenwick~Jr, Klima, McRae, and
  Vahlbusch}{Lee et~al\mbox{.}}{2019}]%
        {lee2019disability}
\bibfield{author}{\bibinfo{person}{Hunter Lee}, \bibinfo{person}{James~B
  Fenwick~Jr}, \bibinfo{person}{Richard~E Klima}, \bibinfo{person}{Alice~A
  McRae}, {and} \bibinfo{person}{Jefford Vahlbusch}.}
  \bibinfo{year}{2019}\natexlab{}.
\newblock \emph{\bibinfo{title}{Disability Assistive Programming: Using Voice
  Input to Write Code}}.
\newblock \bibinfo{thesistype}{Ph.D. Dissertation}.
  \bibinfo{school}{Appalachian State University}.
\newblock


\bibitem[\protect\citeauthoryear{Lei, Wang, Ma, Gan, Lu, Kan, and Chua}{Lei
  et~al\mbox{.}}{2020}]%
        {lei2020re}
\bibfield{author}{\bibinfo{person}{Wenqiang Lei}, \bibinfo{person}{Weixin
  Wang}, \bibinfo{person}{Zhixin Ma}, \bibinfo{person}{Tian Gan},
  \bibinfo{person}{Wei Lu}, \bibinfo{person}{Min-Yen Kan}, {and}
  \bibinfo{person}{Tat-Seng Chua}.} \bibinfo{year}{2020}\natexlab{}.
\newblock \showarticletitle{Re-examining the Role of Schema Linking in
  Text-to-SQL}. In \bibinfo{booktitle}{\emph{Proceedings of the 2020 Conference
  on Empirical Methods in Natural Language Processing (EMNLP)}}.
  \bibinfo{pages}{6943--6954}.
\newblock


\bibitem[\protect\citeauthoryear{Li and Jagadish}{Li and Jagadish}{2014a}]%
        {li2014constructing}
\bibfield{author}{\bibinfo{person}{Fei Li} {and} \bibinfo{person}{HV
  Jagadish}.} \bibinfo{year}{2014}\natexlab{a}.
\newblock \showarticletitle{Constructing an interactive natural language
  interface for relational databases}.
\newblock \bibinfo{journal}{\emph{Proceedings of the VLDB Endowment}}
  \bibinfo{volume}{8}, \bibinfo{number}{1} (\bibinfo{year}{2014}),
  \bibinfo{pages}{73--84}.
\newblock


\bibitem[\protect\citeauthoryear{Li and Jagadish}{Li and Jagadish}{2014b}]%
        {li2014nalir}
\bibfield{author}{\bibinfo{person}{Fei Li} {and} \bibinfo{person}{Hosagrahar~V
  Jagadish}.} \bibinfo{year}{2014}\natexlab{b}.
\newblock \showarticletitle{NaLIR: an interactive natural language interface
  for querying relational databases}. In \bibinfo{booktitle}{\emph{Proceedings
  of the 2014 ACM SIGMOD international conference on Management of data}}.
  \bibinfo{pages}{709--712}.
\newblock


\bibitem[\protect\citeauthoryear{Li, Muller, Thabet, and Ghanem}{Li
  et~al\mbox{.}}{2019}]%
        {li2019deepgcns}
\bibfield{author}{\bibinfo{person}{Guohao Li}, \bibinfo{person}{Matthias
  Muller}, \bibinfo{person}{Ali Thabet}, {and} \bibinfo{person}{Bernard
  Ghanem}.} \bibinfo{year}{2019}\natexlab{}.
\newblock \showarticletitle{Deepgcns: Can gcns go as deep as cnns?}. In
  \bibinfo{booktitle}{\emph{Proceedings of the IEEE/CVF International
  Conference on Computer Vision}}. \bibinfo{pages}{9267--9276}.
\newblock


\bibitem[\protect\citeauthoryear{Li, Zhang, Jia, Xu, Zhang, Wang, Ma, and
  Gao}{Li et~al\mbox{.}}{2020}]%
        {li2020direct}
\bibfield{author}{\bibinfo{person}{Jiguo Li}, \bibinfo{person}{Xinfeng Zhang},
  \bibinfo{person}{Chuanmin Jia}, \bibinfo{person}{Jizheng Xu},
  \bibinfo{person}{Li Zhang}, \bibinfo{person}{Yue Wang},
  \bibinfo{person}{Siwei Ma}, {and} \bibinfo{person}{Wen Gao}.}
  \bibinfo{year}{2020}\natexlab{}.
\newblock \showarticletitle{Direct speech-to-image translation}.
\newblock \bibinfo{journal}{\emph{IEEE Journal of Selected Topics in Signal
  Processing}} \bibinfo{volume}{14}, \bibinfo{number}{3}
  (\bibinfo{year}{2020}), \bibinfo{pages}{517--529}.
\newblock


\bibitem[\protect\citeauthoryear{Luong, Pham, and Manning}{Luong
  et~al\mbox{.}}{2015}]%
        {luong2015effective}
\bibfield{author}{\bibinfo{person}{Thang Luong}, \bibinfo{person}{Hieu Pham},
  {and} \bibinfo{person}{Christopher~D Manning}.}
  \bibinfo{year}{2015}\natexlab{}.
\newblock \showarticletitle{Effective Approaches to Attention-based Neural
  Machine Translation}. In \bibinfo{booktitle}{\emph{EMNLP}}.
\newblock


\bibitem[\protect\citeauthoryear{Lyons, Tran, Binnig, Cetintemel, and
  Kraska}{Lyons et~al\mbox{.}}{2016}]%
        {lyons2016making}
\bibfield{author}{\bibinfo{person}{Gabriel Lyons}, \bibinfo{person}{Vinh Tran},
  \bibinfo{person}{Carsten Binnig}, \bibinfo{person}{Ugur Cetintemel}, {and}
  \bibinfo{person}{Tim Kraska}.} \bibinfo{year}{2016}\natexlab{}.
\newblock \showarticletitle{Making the case for query-by-voice with echoquery}.
  In \bibinfo{booktitle}{\emph{Proceedings of the 2016 International Conference
  on Management of Data}}. \bibinfo{pages}{2129--2132}.
\newblock


\bibitem[\protect\citeauthoryear{Medsker and Jain}{Medsker and Jain}{2001}]%
        {medsker2001recurrent}
\bibfield{author}{\bibinfo{person}{Larry~R Medsker} {and} \bibinfo{person}{LC
  Jain}.} \bibinfo{year}{2001}\natexlab{}.
\newblock \showarticletitle{Recurrent neural networks}.
\newblock \bibinfo{journal}{\emph{Design and Applications}}
  \bibinfo{volume}{5} (\bibinfo{year}{2001}).
\newblock


\bibitem[\protect\citeauthoryear{Nguyen}{Nguyen}{1994}]%
        {nguyen1994near}
\bibfield{author}{\bibinfo{person}{Truong~Q Nguyen}.}
  \bibinfo{year}{1994}\natexlab{}.
\newblock \showarticletitle{Near-perfect-reconstruction pseudo-QMF banks}.
\newblock \bibinfo{journal}{\emph{IEEE Transactions on signal processing}}
  \bibinfo{volume}{42}, \bibinfo{number}{1} (\bibinfo{year}{1994}),
  \bibinfo{pages}{65--76}.
\newblock


\bibitem[\protect\citeauthoryear{Nihalani, Silakari, and Motwani}{Nihalani
  et~al\mbox{.}}{2011}]%
        {nihalani2011natural}
\bibfield{author}{\bibinfo{person}{Neelu Nihalani}, \bibinfo{person}{Sanjay
  Silakari}, {and} \bibinfo{person}{Mahesh Motwani}.}
  \bibinfo{year}{2011}\natexlab{}.
\newblock \showarticletitle{Natural language interface for database: a brief
  review}.
\newblock \bibinfo{journal}{\emph{International Journal of Computer Science
  Issues (IJCSI)}} \bibinfo{volume}{8}, \bibinfo{number}{2}
  (\bibinfo{year}{2011}), \bibinfo{pages}{600}.
\newblock


\bibitem[\protect\citeauthoryear{Nowogrodzki}{Nowogrodzki}{2018}]%
        {nowogrodzki2018speaking}
\bibfield{author}{\bibinfo{person}{Anna Nowogrodzki}.}
  \bibinfo{year}{2018}\natexlab{}.
\newblock \showarticletitle{Speaking in code: how to program by voice}.
\newblock \bibinfo{journal}{\emph{Nature}} \bibinfo{volume}{559},
  \bibinfo{number}{7712} (\bibinfo{year}{2018}), \bibinfo{pages}{141--143}.
\newblock


\bibitem[\protect\citeauthoryear{Obaido, Ade-Ibijola, and Vadapalli}{Obaido
  et~al\mbox{.}}{[n.d.]}]%
        {obaidotalksql}
\bibfield{author}{\bibinfo{person}{George Obaido}, \bibinfo{person}{Abejide
  Ade-Ibijola}, {and} \bibinfo{person}{Hima Vadapalli}.}
  \bibinfo{year}{[n.d.]}\natexlab{}.
\newblock \showarticletitle{TalkSQL: A Tool for the Synthesis of SQL Queries
  from Verbal Specifications}.
\newblock  (\bibinfo{year}{[n.\,d.]}).
\newblock


\bibitem[\protect\citeauthoryear{Peng, Mo, Zhu, Chen, Chen, Xu, and Ma}{Peng
  et~al\mbox{.}}{2020}]%
        {peng2020understanding}
\bibfield{author}{\bibinfo{person}{Zhenhui Peng}, \bibinfo{person}{Kaixiang
  Mo}, \bibinfo{person}{Xiaogang Zhu}, \bibinfo{person}{Junlin Chen},
  \bibinfo{person}{Zhijun Chen}, \bibinfo{person}{Qian Xu}, {and}
  \bibinfo{person}{Xiaojuan Ma}.} \bibinfo{year}{2020}\natexlab{}.
\newblock \showarticletitle{Understanding User Perceptions of Robot's Delay,
  Voice Quality-Speed Trade-off and GUI during Conversation}. In
  \bibinfo{booktitle}{\emph{Extended Abstracts of the 2020 CHI Conference on
  Human Factors in Computing Systems}}. \bibinfo{pages}{1--8}.
\newblock


\bibitem[\protect\citeauthoryear{Povey, Ghoshal, Boulianne, Burget, Glembek,
  Goel, Hannemann, Motlicek, Qian, Schwarz, et~al\mbox{.}}{Povey
  et~al\mbox{.}}{2011}]%
        {povey2011kaldi}
\bibfield{author}{\bibinfo{person}{Daniel Povey}, \bibinfo{person}{Arnab
  Ghoshal}, \bibinfo{person}{Gilles Boulianne}, \bibinfo{person}{Lukas Burget},
  \bibinfo{person}{Ondrej Glembek}, \bibinfo{person}{Nagendra Goel},
  \bibinfo{person}{Mirko Hannemann}, \bibinfo{person}{Petr Motlicek},
  \bibinfo{person}{Yanmin Qian}, \bibinfo{person}{Petr Schwarz},
  {et~al\mbox{.}}} \bibinfo{year}{2011}\natexlab{}.
\newblock \showarticletitle{The Kaldi speech recognition toolkit}. In
  \bibinfo{booktitle}{\emph{IEEE 2011 workshop on automatic speech recognition
  and understanding}}. IEEE Signal Processing Society.
\newblock


\bibitem[\protect\citeauthoryear{Rao, Sak, and Prabhavalkar}{Rao
  et~al\mbox{.}}{2017}]%
        {rao2017exploring}
\bibfield{author}{\bibinfo{person}{Kanishka Rao}, \bibinfo{person}{Ha{\c{s}}im
  Sak}, {and} \bibinfo{person}{Rohit Prabhavalkar}.}
  \bibinfo{year}{2017}\natexlab{}.
\newblock \showarticletitle{Exploring architectures, data and units for
  streaming end-to-end speech recognition with rnn-transducer}. In
  \bibinfo{booktitle}{\emph{2017 IEEE Automatic Speech Recognition and
  Understanding Workshop (ASRU)}}. IEEE, \bibinfo{pages}{193--199}.
\newblock


\bibitem[\protect\citeauthoryear{Ren, Hu, Qin, Zhao, Zhao, and Liu}{Ren
  et~al\mbox{.}}{2020}]%
        {ren2020fastspeech}
\bibfield{author}{\bibinfo{person}{Yi Ren}, \bibinfo{person}{Chenxu Hu},
  \bibinfo{person}{Tao Qin}, \bibinfo{person}{Sheng Zhao},
  \bibinfo{person}{Zhou Zhao}, {and} \bibinfo{person}{Tie-Yan Liu}.}
  \bibinfo{year}{2020}\natexlab{}.
\newblock \showarticletitle{Fastspeech 2: Fast and high-quality end-to-end
  text-to-speech}.
\newblock \bibinfo{journal}{\emph{arXiv preprint arXiv:2006.04558}}
  (\bibinfo{year}{2020}).
\newblock


\bibitem[\protect\citeauthoryear{Sen, Lei, Quamar, {\"O}zcan, Efthymiou,
  Dalmia, Stager, Mittal, Saha, and Sankaranarayanan}{Sen
  et~al\mbox{.}}{2020}]%
        {sen2020athena}
\bibfield{author}{\bibinfo{person}{Jaydeep Sen}, \bibinfo{person}{Chuan Lei},
  \bibinfo{person}{Abdul Quamar}, \bibinfo{person}{Fatma {\"O}zcan},
  \bibinfo{person}{Vasilis Efthymiou}, \bibinfo{person}{Ayushi Dalmia},
  \bibinfo{person}{Greg Stager}, \bibinfo{person}{Ashish Mittal},
  \bibinfo{person}{Diptikalyan Saha}, {and} \bibinfo{person}{Karthik
  Sankaranarayanan}.} \bibinfo{year}{2020}\natexlab{}.
\newblock \showarticletitle{Athena++ natural language querying for complex
  nested sql queries}.
\newblock \bibinfo{journal}{\emph{Proceedings of the VLDB Endowment}}
  \bibinfo{volume}{13}, \bibinfo{number}{12} (\bibinfo{year}{2020}),
  \bibinfo{pages}{2747--2759}.
\newblock


\bibitem[\protect\citeauthoryear{Shah, Li, Kumar, and Saul}{Shah
  et~al\mbox{.}}{2020}]%
        {shah2020speakql}
\bibfield{author}{\bibinfo{person}{Vraj Shah}, \bibinfo{person}{Side Li},
  \bibinfo{person}{Arun Kumar}, {and} \bibinfo{person}{Lawrence Saul}.}
  \bibinfo{year}{2020}\natexlab{}.
\newblock \showarticletitle{SpeakQL: Towards Speech-driven Multimodal Querying
  of Structured Data}. In \bibinfo{booktitle}{\emph{Proceedings of the 2020 ACM
  SIGMOD International Conference on Management of Data}}.
  \bibinfo{pages}{2363--2374}.
\newblock


\bibitem[\protect\citeauthoryear{Shah, Li, Yang, Kumar, and Saul}{Shah
  et~al\mbox{.}}{2019}]%
        {shah2019demonstration}
\bibfield{author}{\bibinfo{person}{Vraj Shah}, \bibinfo{person}{Side Li},
  \bibinfo{person}{Kevin Yang}, \bibinfo{person}{Arun Kumar}, {and}
  \bibinfo{person}{Lawrence Saul}.} \bibinfo{year}{2019}\natexlab{}.
\newblock \showarticletitle{Demonstration of SpeakQL: Speech-driven Multimodal
  Querying of Structured Data}. In \bibinfo{booktitle}{\emph{Proceedings of the
  2019 International Conference on Management of Data}}.
  \bibinfo{pages}{2001--2004}.
\newblock


\bibitem[\protect\citeauthoryear{Shekarpour, Marx, Ngomo, and Auer}{Shekarpour
  et~al\mbox{.}}{2015}]%
        {shekarpour2015sina}
\bibfield{author}{\bibinfo{person}{Saeedeh Shekarpour}, \bibinfo{person}{Edgard
  Marx}, \bibinfo{person}{Axel-Cyrille~Ngonga Ngomo}, {and}
  \bibinfo{person}{S{\"o}ren Auer}.} \bibinfo{year}{2015}\natexlab{}.
\newblock \showarticletitle{Sina: Semantic interpretation of user queries for
  question answering on interlinked data}.
\newblock \bibinfo{journal}{\emph{Journal of Web Semantics}}
  \bibinfo{volume}{30} (\bibinfo{year}{2015}), \bibinfo{pages}{39--51}.
\newblock


\bibitem[\protect\citeauthoryear{Song, Jiang, Huang, Li, Xu, Wong, and
  Yang}{Song et~al\mbox{.}}{2020}]%
        {song2020goldenretriever}
\bibfield{author}{\bibinfo{person}{Yuanfeng Song}, \bibinfo{person}{Di Jiang},
  \bibinfo{person}{Xiaoling Huang}, \bibinfo{person}{Yawen Li},
  \bibinfo{person}{Qian Xu}, \bibinfo{person}{Raymond Chi~Wing Wong}, {and}
  \bibinfo{person}{Qiang Yang}.} \bibinfo{year}{2020}\natexlab{}.
\newblock \showarticletitle{GoldenRetriever: A Speech Recognition System
  Powered by Modern Information Retrieval}. In \bibinfo{booktitle}{\emph{MM}}.
\newblock


\bibitem[\protect\citeauthoryear{Song, Jiang, Zhao, Xu, Wong, Fan, and
  Yang}{Song et~al\mbox{.}}{2021}]%
        {song2021l2rs}
\bibfield{author}{\bibinfo{person}{Yuanfeng Song}, \bibinfo{person}{Di Jiang},
  \bibinfo{person}{Xuefang Zhao}, \bibinfo{person}{Qian Xu},
  \bibinfo{person}{Raymond Chi-Wing Wong}, \bibinfo{person}{Lixin Fan}, {and}
  \bibinfo{person}{Qiang Yang}.} \bibinfo{year}{2021}\natexlab{}.
\newblock \showarticletitle{L2RS: a learning-to-rescore mechanism for automatic
  speech recognition}. In \bibinfo{booktitle}{\emph{MM}}.
\newblock


\bibitem[\protect\citeauthoryear{Sun, Yang, and Liu}{Sun et~al\mbox{.}}{2020}]%
        {sun2020tableqa}
\bibfield{author}{\bibinfo{person}{Ningyuan Sun}, \bibinfo{person}{Xuefeng
  Yang}, {and} \bibinfo{person}{Yunfeng Liu}.} \bibinfo{year}{2020}\natexlab{}.
\newblock \showarticletitle{TableQA: a Large-Scale Chinese Text-to-SQL Dataset
  for Table-Aware SQL Generation}.
\newblock \bibinfo{journal}{\emph{arXiv}} (\bibinfo{year}{2020}),
  \bibinfo{pages}{arXiv--2006}.
\newblock


\bibitem[\protect\citeauthoryear{Sun, Tang, Duan, Ji, Cao, Feng, Qin, Liu, and
  Zhou}{Sun et~al\mbox{.}}{2018}]%
        {sun2018semantic}
\bibfield{author}{\bibinfo{person}{Yibo Sun}, \bibinfo{person}{Duyu Tang},
  \bibinfo{person}{Nan Duan}, \bibinfo{person}{Jianshu Ji},
  \bibinfo{person}{Guihong Cao}, \bibinfo{person}{Xiaocheng Feng},
  \bibinfo{person}{Bing Qin}, \bibinfo{person}{Ting Liu}, {and}
  \bibinfo{person}{Ming Zhou}.} \bibinfo{year}{2018}\natexlab{}.
\newblock \showarticletitle{Semantic parsing with syntax-and table-aware sql
  generation}.
\newblock \bibinfo{journal}{\emph{arXiv preprint arXiv:1804.08338}}
  (\bibinfo{year}{2018}).
\newblock


\bibitem[\protect\citeauthoryear{Tian, Yi, Tao, Bai, and Wen}{Tian
  et~al\mbox{.}}{2019}]%
        {tian2019self}
\bibfield{author}{\bibinfo{person}{Zhengkun Tian}, \bibinfo{person}{Jiangyan
  Yi}, \bibinfo{person}{Jianhua Tao}, \bibinfo{person}{Ye Bai}, {and}
  \bibinfo{person}{Zhengqi Wen}.} \bibinfo{year}{2019}\natexlab{}.
\newblock \showarticletitle{Self-Attention Transducers for End-to-End Speech
  Recognition}.
\newblock \bibinfo{journal}{\emph{Proc. Interspeech 2019}}
  (\bibinfo{year}{2019}), \bibinfo{pages}{4395--4399}.
\newblock


\bibitem[\protect\citeauthoryear{Trummer}{Trummer}{2020}]%
        {trummer2020demonstrating}
\bibfield{author}{\bibinfo{person}{Immanuel Trummer}.}
  \bibinfo{year}{2020}\natexlab{}.
\newblock \showarticletitle{Demonstrating the voice-based exploration of large
  data sets with CiceroDB-zero}.
\newblock \bibinfo{journal}{\emph{Proceedings of the VLDB Endowment}}
  \bibinfo{volume}{13}, \bibinfo{number}{12} (\bibinfo{year}{2020}),
  \bibinfo{pages}{2869--2872}.
\newblock


\bibitem[\protect\citeauthoryear{Utama, Weir, Binnig, and Cetintemel}{Utama
  et~al\mbox{.}}{2017}]%
        {utama2017voice}
\bibfield{author}{\bibinfo{person}{Prasetya Utama}, \bibinfo{person}{Nathaniel
  Weir}, \bibinfo{person}{Carsten Binnig}, {and} \bibinfo{person}{Ugur
  Cetintemel}.} \bibinfo{year}{2017}\natexlab{}.
\newblock \showarticletitle{Voice-based data exploration: Chatting with your
  database}. In \bibinfo{booktitle}{\emph{Proceedings of the Workshop on
  Search-Oriented Conversational AI (SCAI)}}.
\newblock


\bibitem[\protect\citeauthoryear{Vaswani, Shazeer, Parmar, Uszkoreit, Jones,
  Gomez, Kaiser, and Polosukhin}{Vaswani et~al\mbox{.}}{2017}]%
        {vaswani2017attention}
\bibfield{author}{\bibinfo{person}{Ashish Vaswani}, \bibinfo{person}{Noam
  Shazeer}, \bibinfo{person}{Niki Parmar}, \bibinfo{person}{Jakob Uszkoreit},
  \bibinfo{person}{Llion Jones}, \bibinfo{person}{Aidan~N Gomez},
  \bibinfo{person}{{\L}ukasz Kaiser}, {and} \bibinfo{person}{Illia
  Polosukhin}.} \bibinfo{year}{2017}\natexlab{}.
\newblock \showarticletitle{Attention is all you need}. In
  \bibinfo{booktitle}{\emph{Proceedings of the 31st International Conference on
  Neural Information Processing Systems}}. \bibinfo{pages}{6000--6010}.
\newblock


\bibitem[\protect\citeauthoryear{Vinyals, Fortunato, and Jaitly}{Vinyals
  et~al\mbox{.}}{2015}]%
        {vinyals2015pointer}
\bibfield{author}{\bibinfo{person}{Oriol Vinyals}, \bibinfo{person}{Meire
  Fortunato}, {and} \bibinfo{person}{Navdeep Jaitly}.}
  \bibinfo{year}{2015}\natexlab{}.
\newblock \showarticletitle{Pointer Networks}.
\newblock \bibinfo{journal}{\emph{Advances in Neural Information Processing
  Systems}}  \bibinfo{volume}{28} (\bibinfo{year}{2015}),
  \bibinfo{pages}{2692--2700}.
\newblock


\bibitem[\protect\citeauthoryear{Wahlster}{Wahlster}{2013}]%
        {wahlster2013verbmobil}
\bibfield{author}{\bibinfo{person}{Wolfgang Wahlster}.}
  \bibinfo{year}{2013}\natexlab{}.
\newblock \bibinfo{booktitle}{\emph{Verbmobil: foundations of speech-to-speech
  translation}}.
\newblock \bibinfo{publisher}{Springer Science \& Business Media}.
\newblock


\bibitem[\protect\citeauthoryear{Waibel, Hanazawa, Hinton, Shikano, and
  Lang}{Waibel et~al\mbox{.}}{1989}]%
        {waibel1989phoneme}
\bibfield{author}{\bibinfo{person}{Alex Waibel}, \bibinfo{person}{Toshiyuki
  Hanazawa}, \bibinfo{person}{Geoffrey Hinton}, \bibinfo{person}{Kiyohiro
  Shikano}, {and} \bibinfo{person}{Kevin~J Lang}.}
  \bibinfo{year}{1989}\natexlab{}.
\newblock \showarticletitle{Phoneme recognition using time-delay neural
  networks}.
\newblock \bibinfo{journal}{\emph{IEEE transactions on acoustics, speech, and
  signal processing}} \bibinfo{volume}{37}, \bibinfo{number}{3}
  (\bibinfo{year}{1989}), \bibinfo{pages}{328--339}.
\newblock


\bibitem[\protect\citeauthoryear{Wang, Shin, Liu, Polozov, and Richardson}{Wang
  et~al\mbox{.}}{2020b}]%
        {rat-sql}
\bibfield{author}{\bibinfo{person}{Bailin Wang}, \bibinfo{person}{Richard
  Shin}, \bibinfo{person}{Xiaodong Liu}, \bibinfo{person}{Oleksandr Polozov},
  {and} \bibinfo{person}{Matthew Richardson}.}
  \bibinfo{year}{2020}\natexlab{b}.
\newblock \showarticletitle{{RAT-SQL}: Relation-Aware Schema Encoding and
  Linking for Text-to-{SQL} Parsers}. In \bibinfo{booktitle}{\emph{Proceedings
  of ACL}}. \bibinfo{pages}{7567--7578}.
\newblock


\bibitem[\protect\citeauthoryear{Wang, Qiao, Zhu, Hanjalic, and
  Scharenborg}{Wang et~al\mbox{.}}{2020a}]%
        {wang2020s2igan}
\bibfield{author}{\bibinfo{person}{X Wang}, \bibinfo{person}{T Qiao},
  \bibinfo{person}{Jihua Zhu}, \bibinfo{person}{A Hanjalic}, {and}
  \bibinfo{person}{OE Scharenborg}.} \bibinfo{year}{2020}\natexlab{a}.
\newblock \showarticletitle{S2IGAN: Speech-to-Image Generation via Adversarial
  Learning}.
\newblock \bibinfo{journal}{\emph{Proceedings of Interspeech 2020}}
  (\bibinfo{year}{2020}).
\newblock


\bibitem[\protect\citeauthoryear{Wang, Qiao, Zhu, Hanjalic, and
  Scharenborg}{Wang et~al\mbox{.}}{2021}]%
        {wang2021generating}
\bibfield{author}{\bibinfo{person}{Xinsheng Wang}, \bibinfo{person}{Tingting
  Qiao}, \bibinfo{person}{Jihua Zhu}, \bibinfo{person}{Alan Hanjalic}, {and}
  \bibinfo{person}{Odette Scharenborg}.} \bibinfo{year}{2021}\natexlab{}.
\newblock \showarticletitle{Generating Images From Spoken Descriptions}.
\newblock \bibinfo{journal}{\emph{IEEE/ACM Transactions on Audio, Speech, and
  Language Processing}}  \bibinfo{volume}{29} (\bibinfo{year}{2021}),
  \bibinfo{pages}{850--865}.
\newblock


\bibitem[\protect\citeauthoryear{Watanabe, Hori, Kim, Hershey, and
  Hayashi}{Watanabe et~al\mbox{.}}{2017}]%
        {watanabe2017hybrid}
\bibfield{author}{\bibinfo{person}{Shinji Watanabe}, \bibinfo{person}{Takaaki
  Hori}, \bibinfo{person}{Suyoun Kim}, \bibinfo{person}{John~R Hershey}, {and}
  \bibinfo{person}{Tomoki Hayashi}.} \bibinfo{year}{2017}\natexlab{}.
\newblock \showarticletitle{Hybrid CTC/attention architecture for end-to-end
  speech recognition}.
\newblock \bibinfo{journal}{\emph{IEEE Journal of Selected Topics in Signal
  Processing}} \bibinfo{volume}{11}, \bibinfo{number}{8}
  (\bibinfo{year}{2017}), \bibinfo{pages}{1240--1253}.
\newblock


\bibitem[\protect\citeauthoryear{Xu, Liu, and Song}{Xu et~al\mbox{.}}{2017a}]%
        {xu2018sqlnet}
\bibfield{author}{\bibinfo{person}{Xiaojun Xu}, \bibinfo{person}{Chang Liu},
  {and} \bibinfo{person}{Dawn Song}.} \bibinfo{year}{2017}\natexlab{a}.
\newblock \showarticletitle{Sqlnet: Generating structured queries from natural
  language without reinforcement learning}.
\newblock \bibinfo{journal}{\emph{arXiv preprint arXiv:1711.04436}}
  (\bibinfo{year}{2017}).
\newblock


\bibitem[\protect\citeauthoryear{Xu, Liu, and Song}{Xu et~al\mbox{.}}{2017b}]%
        {xu2017sqlnet}
\bibfield{author}{\bibinfo{person}{Xiaojun Xu}, \bibinfo{person}{Chang Liu},
  {and} \bibinfo{person}{Dawn Song}.} \bibinfo{year}{2017}\natexlab{b}.
\newblock \showarticletitle{Sqlnet: Generating structured queries from natural
  language without reinforcement learning}.
\newblock \bibinfo{journal}{\emph{arXiv preprint arXiv:1711.04436}}
  (\bibinfo{year}{2017}).
\newblock


\bibitem[\protect\citeauthoryear{Yu, Li, Zhang, Zhang, and Radev}{Yu
  et~al\mbox{.}}{2018a}]%
        {yu2018typesql}
\bibfield{author}{\bibinfo{person}{Tao Yu}, \bibinfo{person}{Zifan Li},
  \bibinfo{person}{Zilin Zhang}, \bibinfo{person}{Rui Zhang}, {and}
  \bibinfo{person}{Dragomir Radev}.} \bibinfo{year}{2018}\natexlab{a}.
\newblock \showarticletitle{Typesql: Knowledge-based type-aware neural
  text-to-sql generation}.
\newblock \bibinfo{journal}{\emph{arXiv preprint arXiv:1804.09769}}
  (\bibinfo{year}{2018}).
\newblock


\bibitem[\protect\citeauthoryear{Yu, Yasunaga, Yang, Zhang, Wang, Li, and
  Radev}{Yu et~al\mbox{.}}{2018b}]%
        {yu2018syntaxsqlnet}
\bibfield{author}{\bibinfo{person}{Tao Yu}, \bibinfo{person}{Michihiro
  Yasunaga}, \bibinfo{person}{Kai Yang}, \bibinfo{person}{Rui Zhang},
  \bibinfo{person}{Dongxu Wang}, \bibinfo{person}{Zifan Li}, {and}
  \bibinfo{person}{Dragomir Radev}.} \bibinfo{year}{2018}\natexlab{b}.
\newblock \showarticletitle{SyntaxSQLNet: Syntax Tree Networks for Complex and
  Cross-Domain Text-to-SQL Task}. In \bibinfo{booktitle}{\emph{Proceedings of
  the 2018 Conference on Empirical Methods in Natural Language Processing}}.
  \bibinfo{pages}{1653--1663}.
\newblock


\bibitem[\protect\citeauthoryear{Yu, Zhang, Yang, Yasunaga, Wang, Li, Ma, Li,
  Yao, Roman, et~al\mbox{.}}{Yu et~al\mbox{.}}{2018c}]%
        {yu2018spider}
\bibfield{author}{\bibinfo{person}{Tao Yu}, \bibinfo{person}{Rui Zhang},
  \bibinfo{person}{Kai Yang}, \bibinfo{person}{Michihiro Yasunaga},
  \bibinfo{person}{Dongxu Wang}, \bibinfo{person}{Zifan Li},
  \bibinfo{person}{James Ma}, \bibinfo{person}{Irene Li},
  \bibinfo{person}{Qingning Yao}, \bibinfo{person}{Shanelle Roman},
  {et~al\mbox{.}}} \bibinfo{year}{2018}\natexlab{c}.
\newblock \showarticletitle{Spider: A Large-Scale Human-Labeled Dataset for
  Complex and Cross-Domain Semantic Parsing and Text-to-SQL Task}. In
  \bibinfo{booktitle}{\emph{Proceedings of the 2018 Conference on Empirical
  Methods in Natural Language Processing}}. \bibinfo{pages}{3911--3921}.
\newblock


\bibitem[\protect\citeauthoryear{Zenz, Zhou, Minack, Siberski, and Nejdl}{Zenz
  et~al\mbox{.}}{2009}]%
        {zenz2009keywords}
\bibfield{author}{\bibinfo{person}{Gideon Zenz}, \bibinfo{person}{Xuan Zhou},
  \bibinfo{person}{Enrico Minack}, \bibinfo{person}{Wolf Siberski}, {and}
  \bibinfo{person}{Wolfgang Nejdl}.} \bibinfo{year}{2009}\natexlab{}.
\newblock \showarticletitle{From keywords to semantic queries—Incremental
  query construction on the Semantic Web}.
\newblock \bibinfo{journal}{\emph{Journal of Web Semantics}}
  \bibinfo{volume}{7}, \bibinfo{number}{3} (\bibinfo{year}{2009}),
  \bibinfo{pages}{166--176}.
\newblock


\bibitem[\protect\citeauthoryear{Zeyer, Bahar, Irie, Schl{\"u}ter, and
  Ney}{Zeyer et~al\mbox{.}}{2019}]%
        {zeyer2019comparison}
\bibfield{author}{\bibinfo{person}{Albert Zeyer}, \bibinfo{person}{Parnia
  Bahar}, \bibinfo{person}{Kazuki Irie}, \bibinfo{person}{Ralf Schl{\"u}ter},
  {and} \bibinfo{person}{Hermann Ney}.} \bibinfo{year}{2019}\natexlab{}.
\newblock \showarticletitle{A comparison of Transformer and LSTM encoder
  decoder models for ASR}. In \bibinfo{booktitle}{\emph{2019 IEEE Automatic
  Speech Recognition and Understanding Workshop (ASRU)}}. IEEE,
  \bibinfo{pages}{8--15}.
\newblock


\bibitem[\protect\citeauthoryear{Zhang, Yu, Er, Shim, Xue, Lin, Shi, Xiong,
  Socher, and Radev}{Zhang et~al\mbox{.}}{2019}]%
        {zhang2019editing}
\bibfield{author}{\bibinfo{person}{Rui Zhang}, \bibinfo{person}{Tao Yu},
  \bibinfo{person}{Heyang Er}, \bibinfo{person}{Sungrok Shim},
  \bibinfo{person}{Eric Xue}, \bibinfo{person}{Xi~Victoria Lin},
  \bibinfo{person}{Tianze Shi}, \bibinfo{person}{Caiming Xiong},
  \bibinfo{person}{Richard Socher}, {and} \bibinfo{person}{Dragomir Radev}.}
  \bibinfo{year}{2019}\natexlab{}.
\newblock \showarticletitle{Editing-Based SQL Query Generation for Cross-Domain
  Context-Dependent Questions}. In \bibinfo{booktitle}{\emph{Proceedings of the
  2019 Conference on Empirical Methods in Natural Language Processing and the
  9th International Joint Conference on Natural Language Processing
  (EMNLP-IJCNLP)}}. \bibinfo{pages}{5338--5349}.
\newblock


\bibitem[\protect\citeauthoryear{Zhao, Wang, He, Yang, Chang, and Wang}{Zhao
  et~al\mbox{.}}{2020}]%
        {zhao2020multiple}
\bibfield{author}{\bibinfo{person}{Xiangyu Zhao}, \bibinfo{person}{Longbiao
  Wang}, \bibinfo{person}{Ruifang He}, \bibinfo{person}{Ting Yang},
  \bibinfo{person}{Jinxin Chang}, {and} \bibinfo{person}{Ruifang Wang}.}
  \bibinfo{year}{2020}\natexlab{}.
\newblock \showarticletitle{Multiple knowledge syncretic transformer for
  natural dialogue generation}. In \bibinfo{booktitle}{\emph{Proceedings of The
  Web Conference 2020}}. \bibinfo{pages}{752--762}.
\newblock


\bibitem[\protect\citeauthoryear{Zheng, Cheng, Zou, Yu, and Zhao}{Zheng
  et~al\mbox{.}}{2017}]%
        {zheng2017natural}
\bibfield{author}{\bibinfo{person}{Weiguo Zheng}, \bibinfo{person}{Hong Cheng},
  \bibinfo{person}{Lei Zou}, \bibinfo{person}{Jeffrey~Xu Yu}, {and}
  \bibinfo{person}{Kangfei Zhao}.} \bibinfo{year}{2017}\natexlab{}.
\newblock \showarticletitle{Natural language question/answering: Let users talk
  with the knowledge graph}. In \bibinfo{booktitle}{\emph{Proceedings of the
  2017 ACM on Conference on Information and Knowledge Management}}.
  \bibinfo{pages}{217--226}.
\newblock


\bibitem[\protect\citeauthoryear{Zhong, Xiong, and Socher}{Zhong
  et~al\mbox{.}}{2017}]%
        {zhong2017seq2sql}
\bibfield{author}{\bibinfo{person}{Victor Zhong}, \bibinfo{person}{Caiming
  Xiong}, {and} \bibinfo{person}{Richard Socher}.}
  \bibinfo{year}{2017}\natexlab{}.
\newblock \showarticletitle{Seq2sql: Generating structured queries from natural
  language using reinforcement learning}.
\newblock \bibinfo{journal}{\emph{arXiv preprint arXiv:1709.00103}}
  (\bibinfo{year}{2017}).
\newblock


\bibitem[\protect\citeauthoryear{Zhou, Dong, Xu, and Xu}{Zhou
  et~al\mbox{.}}{2018}]%
        {zhou2018comparison}
\bibfield{author}{\bibinfo{person}{Shiyu Zhou}, \bibinfo{person}{Linhao Dong},
  \bibinfo{person}{Shuang Xu}, {and} \bibinfo{person}{Bo Xu}.}
  \bibinfo{year}{2018}\natexlab{}.
\newblock \showarticletitle{A comparison of modeling units in
  sequence-to-sequence speech recognition with the transformer on mandarin
  chinese}. In \bibinfo{booktitle}{\emph{International Conference on Neural
  Information Processing}}. Springer, \bibinfo{pages}{210--220}.
\newblock


\end{thebibliography}

\end{document}